\documentclass[12pt,preprint,tightenlines,
showpacs,preprintnumbers,amsmath,amssymb,
a4paper,nofootinbib]{revtex4-2}

\usepackage{nicefrac}

\usepackage{natbib}
\usepackage{amssymb,amsbsy,amsmath,amsfonts}
\usepackage{graphicx}
\usepackage{epsf,epsfig,float,latexsym,amsthm,fancyhdr,rotating}
\usepackage{graphics,psfrag,longtable}
\usepackage{slashed}
\usepackage[colorlinks,citecolor=blue,linktoc=all,linkcolor=cyan,urlcolor=blue]{hyperref}
\usepackage{upgreek}

\def\beq{\begin{equation}}
\def\eeq{\end{equation}}
\def\bea{\begin{eqnarray}}
\def\eea{\end{eqnarray}}
\def\beqa{\begin{equation}\begin{array}{l}}
\def\eeqa{\end{array}\end{equation}}
\def\eqlab#1{\label{eq:#1}}

\def\seclab#1{\label{sec:#1}}
\def\eref#1{(\ref{eq:#1})}
\def\Eqref#1{Eq.~(\ref{eq:#1})}

\def\Figref#1{Fig.~\ref{fig:#1}}

\def\secref#1{Section \ref{sec:#1}}

\def\barr{\left(\begin{array}{c}}
\def\earr{\end{array}\right)}
\def\bmat{\left(\begin{array}{cc}}
\def\emat{\end{array}\right)}
\def\al{\alpha}
\def\be{\beta}
\def\ga{\gamma} 
\def\de{\delta} \def\De{\Delta}

 \def\La{{\Lambda}}

\def\si{\sigma}

\def\pa{\partial}

\def\pa{\partial}

\def\nn{\nonumber}
\def\dd{\mathrm{d}}

\def\lag{{\mathcal L}}

\DeclareMathOperator\im{Im}
\def\3d{3-D}

\def\ol#1{\overline{#1}}

\begin{document}
\title {Forward doubly-virtual 
Compton scattering off the nucleon in chiral perturbation theory:
the subtraction function and moments of unpolarized structure functions}
\author{Jose Manuel Alarc\'on}
\affiliation{Departamento de F\'isica Te\'orica \& IPARCOS, Universidad Complutense de Madrid, 28040
Madrid, Spain}
\author{Franziska Hagelstein}
\affiliation{Albert Einstein Center for Fundamental Physics, Institute for Theoretical Physics, University of Bern, Sidlerstrasse 5, CH-3012 Bern, Switzerland}
\author{Vadim Lensky}
\affiliation{Institut f\"ur Kernphysik \&  Cluster of Excellence PRISMA+,
 Johannes Gutenberg-Universit\"at  Mainz,  D-55128 Mainz, Germany}
\author{Vladimir Pascalutsa}
\affiliation{Institut f\"ur Kernphysik \&  Cluster of Excellence PRISMA+,
 Johannes Gutenberg-Universit\"at  Mainz,  D-55128 Mainz, Germany}
 \email{vladipas@kph.uni-mainz.de}
\begin{abstract}
The forward doubly-virtual 
Compton scattering (VVCS) off the nucleon contains a wealth of information on nucleon structure, relevant to the calculation
of the two-photon-exchange effects in atomic spectroscopy and electron scattering. We report on a complete next-to-leading-order (NLO)
calculation of low-energy VVCS in chiral perturbation
theory ($\chi$PT). 
Here we focus on the unpolarized VVCS amplitudes $T_1(\nu, Q^2)$
and $T_2(\nu, Q^2)$, and the corresponding structure functions
$F_1(x, Q^2)$ and $F_2(x,Q^2)$. Our results are confronted, where possible, with ``data-driven'' dispersive evaluations of 
low-energy structure 
quantities, such as nucleon polarizabilities.  We find  significant disagreements with dispersive evaluations at very
low momentum-transfer $Q$; for example,
in the slope of polarizabilities at zero momentum-transfer. 
By expanding the results in powers of the inverse nucleon  mass, we reproduce the known ``heavy-baryon'' expressions. This serves as a check of our calculation, as well as  demonstrates the differences between the manifestly Lorentz-invariant (B$\chi$PT)
and heavy-baryon (HB$\chi$PT) frameworks. 
\end{abstract}
\date{\today}
\preprint{MITP/20-032}
\maketitle
\newpage

\tableofcontents

\newpage
\section{Introduction and outline}

\begin{figure}[tbh]
\centering
       \includegraphics[width=5.5cm]{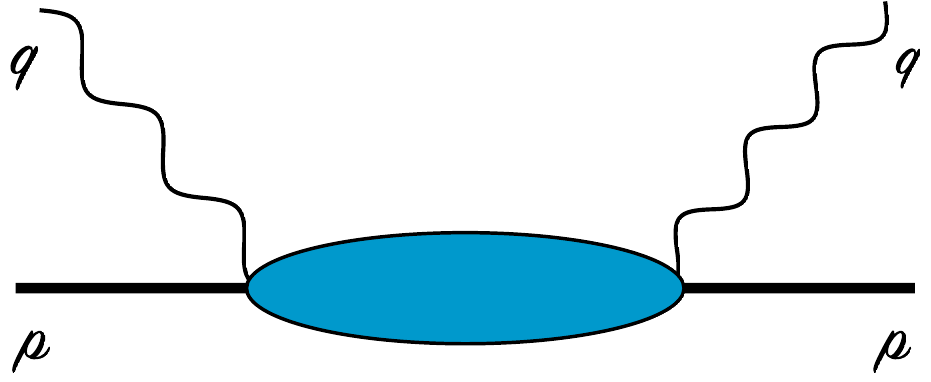}
\caption{\small{The forward Compton scattering, or VVCS, in case of virtual photons, $q^2=-Q^2$.
  \label{fig:CSgeneric}}}
\end{figure}

The forward doubly-virtual 
Compton scattering (VVCS), \Figref{CSgeneric}, is not a directly observable process. Nevertheless, it 
is traditionally of high relevance in studies
of nucleon and nuclear structure, and of their impact on atomic nuclei. At high energies
the VVCS has the  apparent connections to deep-inelastic scattering, whereas 
at low energies it is important for precision atomic spectroscopy, where it serves as input for calculations of the nuclear-structure corrections. Analytical properties of the VVCS amplitude
are used to establish useful relations --- sum rules ---
between the static 
(electromagnetic moments, polarizabilities) 
and dynamic (photoabsorption cross sections) properties of the nucleon \cite{GellMann:1954db,Burkhardt:1970ti,Schwinger:1975uq,Bernabeu:1974mb,Lvov:1998vg},
see also Refs.~\cite{Drechsel:2002ar,Kuhn:2008sy,Holstein:2013kia,Hagelstein:2015egb,Pascalutsa:2018ced,Pasquini:2018wbl} for reviews. 

In the past decade, with the advent of muonic-atom spectroscopy
by the CREMA Collaboration \cite{Pohl:2010zza,Antognini:1900ns,Pohl1:2016xoo},
the interest in nucleon VVCS has resurged in the context of 
the  ``proton radius puzzle'' (see, e.g., Refs.~\cite{Pohl:2013yb,Carlson:2015jba} for reviews).  
The muonic atoms, being more sensitive to nuclear structure than
conventional atoms, demand a higher quality of this input
in both the Lamb shift \cite{Pohl:2010zza,Antognini:1900ns} and, in the near future, the hyperfine structure measurements  \cite{Pohl:2016xsr,Bakalov:2015xya,Kanda:2018oay}. The VVCS  enters here in the form of the two-photon exchange (TPE) corrections appearing at $\mathcal{O}(Z^4\al^5)$, which is the subleading order for the nuclear-structure effects in the Lamb shift (the leading  being the  charge radius), and leading in the hyperfine structure. In either case, the TPE is the leading theoretical uncertainty and precising this contribution is a challenge for the nuclear and hadron
physics community.

In this work we focus on the unpolarized nucleon 
VVCS, described, for each nucleon (proton or neutron), by two scalar amplitudes $T_{1,2}(\nu,Q^2)$,
functions of the photon energy $\nu$ and virtuality $Q^2$.
The  discontinuity of these amplitudes is
given, respectively, by the two unpolarized structure functions $F_1(x,Q^2)$ and
$F_{2}(x, Q^2)$.

To date, there are two approaches: 
\textit{1) dispersion relations (DR)} and 
\textit{2) chiral perturbation theory ($
\chi$PT)}, used for evaluation
of nucleon VVCS, with the goal of quantifying the relevant corrections in muonic hydrogen.
It is expected and  highly desirable that 
\textit{3) lattice QCD} will join this effort in the near future. 
In the mean time, however, the DR approach is the most popular one.
It employs the well-known dispersion relations expressing 
the VVCS amplitudes as integrals of the structure
functions known empirically from inclusive electron scattering.

Unfortunately, the DRs determine the VVCS in terms of the structure functions only up to a ``subtraction  function'' $T_1(0,Q^2)$. 
The latter function is not well-constrained empirically, which
makes this approach prone to model uncertainties. It is worthwhile to mention that there is a
new proposal on how the subtraction can further be constrained
via the dilepton electroproduction \cite{Pauk:2020gjv}. However, in the foreseeable future,
this issue will preclude a systematic improvement of the theoretical uncertainty within the DR approach.

Here we employ the second approach. More specifically, we use an extension of SU(2) $\chi$PT \cite{Weinberg:1978kz,Gasser:1983yg} to the single-baryon sector \cite{Gasser:1987rb,Gegelia:1999gf,Fuchs:2003qc}, referred to as the baryon $\chi$PT (B$\chi$PT), augmented by inclusion 
of the explicit $\Delta(1232)$-isobar in the $\de$-counting scheme
\cite{Pascalutsa:2003aa}. 
In this framework we compute the inelastic (non-Born) part of the
VVCS amplitudes to next-to-leading order (NLO).
A first version of this calculation was briefly considered in~Ref.~\cite{Lensky:2014dda}. Here we provide a few important improvements, in particular, the inclusion of the Coulomb-quadrupole $(C2)$ $N\to \Delta$ transition, and a more comprehensive comparison of our results with the DR approach.
The impact of this
calculation on the muonic-hydrogen Lamb shift, extending
our previous evaluation~\cite{Alarcon:2013cba} to higher orders, will be discussed elsewhere.

The paper is organized as follows. In Sec.~\ref{sec:LorentzDecomp}, we recall the general formulas for VVCS and its relation to structure functions, form factors and polarizabilities. In Sec.~\ref{Sec:ChEFT_for_VVCS},  we discuss
the main ingredients of our NLO calculation. 
In Sec.~\ref{Sec:Scalar-Pol}, we examine results for the proton and neutron scalar polarizabilities, and some of the other moments
of structure functions.  In the concluding section 
(Sec.\ \ref{Sec:Summary}), we summarize and give a brief outlook
for the near-future work. In App.~\ref{App:CrossSections}, we discuss the structure functions, in particular, the $\pi N$, $\pi \De$ and $\De$ production channels  
relevant to our calculation. In App.~\ref{App:PolarizabilitiesAll}, 
we give analytical expressions for the $\pi N$-loop and $\Delta$-exchange contributions to the central values and slopes of the polarizabilities and moments of structure functions at $Q^2=0$. The complete expressions, also for the $\pi \Delta$-loop contributions, can be found in the {\it Supplemental Material}.

\section{VVCS formalism} \seclab{LorentzDecomp}

Figure \ref{fig:CSgeneric} schematically shows  the VVCS 
amplitude, which for an unpolarized target  (of any spin) can be decomposed into two
independent Lorentz-covariant and gauge-invariant tensor structures
\cite{Hagelstein:2015egb}:
\bea
\label{Eq:T-Rel}
T^{\mu\nu}(p,q) & = &  
\left( -g^{\mu\nu}+\frac{q^{\mu}q^{\nu}}{q^2}\right)
T_1(\nu, Q^2) +\frac{1}{M_N^2} \left(p^{\mu}-\frac{p\cdot
q}{q^2}\,q^{\mu}\right) \left(p^{\nu}-\frac{p\cdot
q}{q^2}\, q^{\nu} \right) T_2(\nu, Q^2),\quad
\eqlab{fVVCS}
\eea
where $p$ and $q$ are the four-momenta of the target particle and the photon, respectively;
$M_N$ is the target (here, nucleon) mass. 
The scalar amplitudes $T_i$ are functions of the photon
energy $\nu = p\cdot q/ M_N$ and virtuality $Q^2=-q^2$.

The optical theorem relates the absorptive parts of the VVCS amplitudes to the structure functions, or equivalently, the inclusive electroproduction cross sections:
\begin{subequations}
\eqlab{VVCSunitarity}
\bea
\im T_1(\nu,Q^2)&=&\frac{4\pi^2\al}{M_N}F_1(x,Q^2) \eqlab{ImT1}\\  &=& K(\nu,Q^2)\,\sigma_T(\nu,Q^2), \nn\\
\im T_2(\nu,Q^2)&=&\frac{4\pi^2\al}{\nu}F_2(x,Q^2)\eqlab{ImT2}\\
 &=& \frac{Q^2  K(\nu,Q^2) }{\nu^2+Q^2}\left[\sigma_T(\nu,Q^2)+\sigma_L(\nu,Q^2)\right],\nn
\eea
\end{subequations}
with the fine-structure constant $\alpha=e^2/4\pi$, and the Bjorken variable $x=Q^2/2M_N \nu$. The two response functions $\sigma_T$ and $\sigma_L$ are cross sections of total photoabsorption  of transversely ($T$) and longitudinally ($L$) polarized photons.
The flux of virtual photons is
conventionally defined up to the flux factor $K(\nu,Q^2)$. The experimental observables do not depend on it, only the definitions of the response functions $\sigma_T$ and $\sigma_L$ do.  
Throughout this work we adopt  Gilman's flux factor 
(for other common choices, cf.~Ref.~\cite{Drechsel:2002ar}):
\beq 
K(\nu,Q^2)=\vert\vec{q}\,\vert=\sqrt{\nu^2+Q^2},
\eeq
where  $\vec{q}$ is the photon three-momentum in the lab frame.

The VVCS amplitudes satisfy the following dispersion relations derived from the above statement of the  optical theorem, combined with general principles of analyticity and crossing symmetry
(cf., for example, Refs.~\cite{Drechsel:2002ar,Hagelstein:2015egb,Pascalutsa:2018ced} for  details):
\begin{subequations}
\eqlab{genDRs}
\bea 
T_1 ( \nu, Q^2) &=&T_1(0,Q^2) + \frac{32\pi\al M_N\nu^2}{Q^4} \int_{0}^1 
\!\dd x \,
\frac{x\, F_1 (x, Q^2)}{1 - x^2 (\nu/\nu_{\mathrm{el}})^2 - i 0^+} \eqlab{T1dr}\\ 
&=& T_1(0,Q^2) + \frac{2\nu^2}{\pi } \int_{\nu_{\mathrm{el}}}^\infty\! \frac{\dd \nu'}{\nu'} \, 
\frac{\sqrt{\nu^{\prime\,2}+Q^2}\,\si_T ( \nu', Q^2)}{\nu^{\prime\,2} -\nu^2 - i 0^+}\,,\nn\\
T_2 ( \nu, Q^2) &=& \frac{16\pi\al M_N}{Q^2} \int_{0}^1 
\!\dd x\, 
\frac{F_2 (x, Q^2)}{1 - x^2 (\nu/\nu_{\mathrm{el}})^2  - i 0^+} \eqlab{T2dr}\\
&=&\frac{2Q^2}{\pi} \int_{\nu_{\mathrm{el}}}^\infty\! \dd \nu' \, 
\frac{\nu^{\prime}\,[ \si_T+\si_L] ( \nu', Q^2)}{\sqrt{\nu^{\prime\,2}+Q^2}(\nu^{\prime\,2} -\nu^2- i 0^+)} ,\nn
\eea 
\end{subequations}
with $\nu_{\mathrm{el}}=Q^2/2M_N$ the elastic threshold. The high-energy behavior of $F_1(x,Q^2)$ prevents the convergence of the corresponding unsubtracted dispersion integral,  hence
leading to the once-subtracted dispersion relation, \Eqref{T1dr}, with the aforementioned ``subtraction function'' $T_1(0,Q^2)$. 
Note that while the subtraction point is conventionally
chosen at $\nu = 0$, 
other choices are in principle possible. Future lattice QCD calculations of the VVCS amplitude would likely  prefer to deal
with a Euclidean subtraction point,  e.g., at $\nu =iQ/2$, as chosen in Ref.~\cite{Gasser:2020mzy}.

The amplitudes are naturally split into nucleon-pole ($T_i^{\mathrm{pole}}$) and non-pole ($T_i^{\mathrm{nonpole}}$)
parts, or Born ($T_i^{\mathrm{Born}}$) and non-Born ($\ol T_i$) terms,
\beq
T_i = T_i^{\mathrm{pole}} + T_i^{\mathrm{nonpole}}
= T_i^{\mathrm{Born}} + \ol T_i,
\eeq 
with the pole and Born terms given uniquely in terms
of the nucleon electric ($G_E$) and magnetic ($G_M$) Sachs form factors:
\begin{subequations}
\bea
T_1^{\mathrm{pole}}   (\nu,Q^2) &=& \frac{4\pi \al}{M_N}
\frac{\nu_\mathrm{el}^2}{\nu_\mathrm{el}^2-\nu^2-i0^+}\, G_M^2(Q^2), \\
 T_2^{\mathrm{pole}}(\nu,Q^2) &=& \frac{8\pi \al\nu_\mathrm{el}}{\nu_\mathrm{el}^2-\nu^2-i0^+}\, \frac{G_E^2(Q^2)+\tau G_M^2(Q^2)}{1+\tau}, \\
T_1^{\mathrm{Born}}(\nu, Q^2) &=& -\frac{4\pi \alpha}{M_N}\left[\frac{G_E(Q^2)+\tau G_M (Q^2)}{1+\tau}\right]^2+T_1^{\mathrm{pole}}(\nu, Q^2),\\ 
T_2^{\mathrm{Born}}(\nu, Q^2) &=& T_2^{\mathrm{pole}}(\nu, Q^2),
\eea
\end{subequations}
where $\tau = Q^2/4 M_N^2$. The $i0^+$ prescription represents the fact that the imaginary part of these amplitudes is given by
the elastic piece of the structure functions: $F_i^{\mathrm{el}}(Q^2) = F_i(x=1, \, Q^2)$. One can thus
exclude the pole piece from the above dispersion relations
by setting the lower-energy limit of integration to an inelastic threshold $\nu_0$ instead of $\nu_\mathrm{el}$, 
or $x_0 = Q^2/2M_N\nu_0$ instead of 1. 
For the nucleon the first inelastic
threshold is usually associated with one-pion production, i.e.,
$\nu_0 = \nu_\mathrm{el}+ m_\pi (1 + m_\pi/2M_N) $, where $m_\pi$ is the pion mass.

We are not concerned here with the elastic form factors, and therefore in the rest of the paper we focus on the non-Born part
of the amplitudes,  $\ol T_i$. The low-energy and low-momentum expansion
of these amplitudes is given in terms of the static polarizabilities, e.g., for the lowest-order terms one obtains
\begin{subequations}
\bea 
\ol T_1(\nu, Q^2)/4\pi  &=& \beta_{M1} Q^2 + (\alpha_{E1} + \beta_{M1}) \nu^2 + \ldots, \\
\ol T_2(\nu, Q^2)/4\pi  &=&  (\alpha_{E1} + \beta_{M1}) Q^2 + \ldots,
\eea 
\end{subequations}
where $\alpha_{E1}$ ($\beta_{M1}$) is  the  electric (magnetic) dipole
polarizability.
Such an expansion of both sides of the dispersion relations \eref{genDRs} thus results in various sum rules, most notably, the Baldin sum rule \cite{Baldin:1960} for 
 $\alpha_{E1} + \beta_{M1}$. Further relations derived from unpolarized VVCS are considered in Ref.~\cite{Lensky:2017bwi}.
 
 More generally, one may expand the dispersion relations \eref{genDRs} in $\nu$ alone, keeping $Q^2$ fixed. On the right-hand side, one finds the moments of structure functions. Introducing
 \begin{subequations}
 \bea
  M_1^{(n)}(Q^2) &\equiv &  \frac{4\al}{Q^2} \left(\frac{2M_N}{Q^2}\right)^{n-1} \int_{0}^{x_0} 
\!\dd x \, x^{n-1} \, F_1(x,Q^2), \\
 M_2^{(n)}(Q^2) &\equiv &  \frac{4\al M_N}{Q^4} \left(\frac{2M_N}{Q^2}\right)^{n-1} \int_{0}^{x_0} 
\!\dd x \, x^{n-1} \, F_2(x,Q^2), 
 \eea 
 \end{subequations}
 the relations \eref{genDRs} lead to:
 \begin{subequations}
 \eqlab{VVCSexpansion}
 \bea
 \ol  T_1(\nu,Q^2) &=& \ol T_1(0,Q^2) + 4\pi \sum_{k=1} M_1^{(2k)}(Q^2)\,  \nu^{2k},\\
  \ol T_2(\nu,Q^2) &=&  4\pi \sum_{k=0} M_2^{(2k+1)}(Q^2)\, \nu^{2k}\,.
 \eea 
 \end{subequations}
 Note that in the limit of $Q^2\to 0$, we obtain the Baldin sum rule
 in the form:
 \beq 
 \alpha_{E1} + \beta_{M1} = M_1^{(2)}(0) =M_2^{(1)}(0).
 \eeq 
 We refer to $M_1^{(2)}(Q^2)$ as the generalized Baldin sum rule, see Sec.~\ref{M12genBS}. 
 More generally, we have the following relation (for an integer $n$):
 \beq
 M_1^{(n)}(0) =M_2^{(n-1)}(0)\, ,
 \eeq 
 arising from electromagnetic gauge invariance. One way to derive it
 is to introduce the longitudinal amplitude
 \beq 
 T_L(\nu,Q^2) = -T_1(\nu, Q^2) +\frac{Q^2+\nu^2}{Q^2} T_2(\nu,Q^2)
 \eeq
 and to show that $\lim_{Q^2\to 0} T_L(\nu,Q^2) =0$. Incidentally,
 the same is true for asymptotically large $Q^2$, because of the
 Callan-Gross relation: $2x F_1(x,Q^2) = F_2(x,Q^2)$, and
 hence $M_1^{(n)}(\infty) =M_2^{(n-1)}(\infty)$.

 In what follows, we consider some of these
 moments, obtaining them from the $\chi$PT results for the VVCS amplitudes, and compare them
 with the results of  empirical parametrizations of the nucleon structure functions.
 The dispersion relations \eref{genDRs} are used by us to cross-check the results, using
 the tree-level photoabsorption cross sections discussed in App.~\ref{App:CrossSections}.

\section{Calculation of the VVCS amplitude at NLO} 
\label{Sec:ChEFT_for_VVCS}

Our goal here is to obtain the $\chi$PT predictions of
the non-Born parts of the nucleon VVCS amplitudes $T_{1,2}$.
The present NLO calculation is still within the ``predictive powers'' of $\chi$PT for Compton scattering (CS) amplitudes, i.e., 
the results are given in terms of well-known parameters
(see Table \ref{tab:constants}) obtained from non-Compton processes.
In this sense, it is complementary to the existing calculations of the real CS (RCS) \cite{Lensky:2009uv,Lensky:2015awa}
and the virtual CS (VCS) \cite{Lensky:2016nui}. All of these studies, 
including the present one, are done in the same framework, using the same set of parameters.

\subsection{Remarks on power counting} \seclab{PowerCounting}

  We shall employ B$\chi$PT, which is the manifestly covariant extension of $\chi$PT to the single-baryon sector in its most straightforward implementation, where
the nucleon is included as in Ref.~\cite{Gasser:1987rb}. The power-counting concerns  
raised in Ref.~\cite{Gasser:1987rb} have been overcome by renormalizing away the ``power-counting
violation'' using the low-energy constants (LECs)
available at that order.
 This has been shown explicitly within the  ``extended on-mass-shell renormalization scheme'' (EOMS)~\cite{Fuchs:2003qc}, 
 but is not limited to it. 
 The inclusion of the explicit $\De(1232)$ here will follow
 the ``$\de$-counting'' framework 
of Ref.~\cite{Pascalutsa:2003aa} (see also Refs.~\cite{Pascalutsa:2006up,Geng:2013xn} for concise overviews).

To explain the power counting in more detail,  
let us recall that  chiral effective-field theory is based on 
a perturbative expansion in powers of pion momentum $p$ and mass $m_\pi$ over the scale of spontaneous chiral symmetry breaking ${\Lambda_\chi\sim 4\pi f_\pi}$, with
$f_\pi\simeq 92$~MeV the pion decay constant. Each operator in the effective Lagrangian, or a
graph in the loopwise expansion of the $S$-matrix, can have a specific order of
$p$ assigned to it.

To give a relevant example consider the following operator: 
\beq
\lag^{(4)} \sim \de \be \, \bar N N F^2,
\eeq
with $\de \be$ the coupling constant, $N$ the Dirac field
of the nucleon, and $F^2$ the square of the electromagnetic field strength tensor, $F_{\mu\nu}=\pa_{[\mu}A_{\nu]}$. This is an operator of $\mathcal{O}(p^4)$. Two of the $p$'s come from the
photon momenta which are supposed to be small, and the other two powers arise because the two-photon
coupling to the nucleon must carry a factor of $\al$ (the charge $e$ counts as $p$, since we want
the derivative of the pion field to count as $p$ even after including the minimal coupling to the photon).

This operator enters the effective Lagrangian with an LEC, which we denote $\de \be$.
It gives a contribution to the CS amplitude in the form of\footnote{
Throughout this paper we use the conventions summarized at the beginning of Ref.~\cite{Hagelstein:2015egb}.}
\beq 
T^{\mu\nu} = 4\pi \, \de \be \, (q\cdot q' \, g^{\mu\nu} - q^\mu q^{\prime\,\nu}),
\eqlab{eq:deltabetaamp}
\eeq 
and leads to a shift in the magnetic dipole polarizability as: $\be_{M1}\to \be_{M1}+\de\be $.
Now, two remarks are in order.
\begin{itemize}
\item[i)] {\em Naturalness}. The magnitude of the LEC is not arbitrary. It goes as 
${\de \be = (\al/\La_\chi^3) c} $, with the dimensionless constant $c$ being of the order of 1, or more precisely:
\beq 
p/\La_\chi \ll |c|\ll \La_\chi/p\,.
\eeq 
This condition ensures that the contribution  of this operator is indeed 
of $\mathcal{O}(p^4)$, as inferred by the power counting.
\item[ii)] {\em Predictive powers}.
This LEC enters very prominently in the polarizabilities and CS  at tree level,
which means its value is best fixed by the empirical information on these quantities. 
If this is so, the $\mathcal{O}(p^4)$ result is not ``predictive'', as it could only be used to {\em fit} 
the $\chi$PT expression to experiment or lattice QCD calculations. On the other hand, contributions
of orders lower than $p^4$ are predictive, as they only contain LECs fixed from elsewhere.
\end{itemize}

As already mentioned, the ``predictive'' contributions to CS and polarizabilities have been identified  and computed  for the case
of RCS \cite{Lensky:2009uv}, 
VCS \cite{Lensky:2016nui}, and  VVCS \cite{Lensky:2014dda}.
Our present calculation is quite analogous to those works and hence 
we refer to them for most of the technical details, such as the expressions for the
relevant terms of the effective Lagrangian. It is crucial to first study these predictive contributions. We note, however,
that here we choose to also include the $p^4$ LEC that shifts the magnetic polarizability. In doing so,
we fit the value of $\delta\beta$ so as to reproduce the Baldin sum rule values:
\begin{subequations}
\eqlab{fittedBaldin}
\begin{align}
\alpha_{E1p}+\beta_{M1p} & = 14.0(0.2)\times 10^{-4} \text{ fm}^3 \text{\qquad proton~\cite{Gryniuk:2015aa},}\\
\alpha_{E1n}+\beta_{M1n} & = 15.2(0.5)\times 10^{-4} \text{ fm}^3 \text{ \qquad neutron~\cite{Levchuk:1999zy}}\,,
\end{align}
\end{subequations}
taking the values of $\alpha_{E1}$ obtained at $\mathcal{O}(p^4/\varDelta)$ as B$\chi$PT predictions.
This choice reflects the fact that the most prominent scalar moments considered here, the second moment
of $F_1(x,Q^2)$ and the first moment of $F_2(x,Q^2)$, both change into the Baldin sum rule in the real-photon limit. The values of the magnetic polarizabilities that result from this fit are
\begin{align}\eqlab{newBeta}
    \beta_{M1p} = 2.75(0.2)\times10^{-4}\text{ fm}^3\,, &\qquad \beta_{M1n} = 1.5(0.5)\times10^{-4}\text{ fm}^3\,,
\end{align}
where the error bar does not include the theoretical uncertainty. One has to admit that this procedure results in somewhat smaller values of $\beta_{M1}$ than, for instance, those obtained in the recent heavy-baryon (HB) and covariant chiral analyses: $3.2(0.5)\times10^{-4}\text{ fm}^3$~\cite{Lensky:2014efa,Griesshammer:2012we} for the proton and $3.65(1.25)\times10^{-4}\text{ fm}^3$~\cite{Myers14} for the neutron.
 We will, however, use this simplified procedure since the only affected quantity studied by us
is the proton subtraction function $\ol{T}_{1p}(0,Q^2)/4\pi=\beta_{M1p}\,Q^2+\dots\,,$ and the
discrepancy for $\beta_{M1p}$ is tolerable.

\begin{figure}[tbh]
\centering
 \includegraphics[width=5cm]{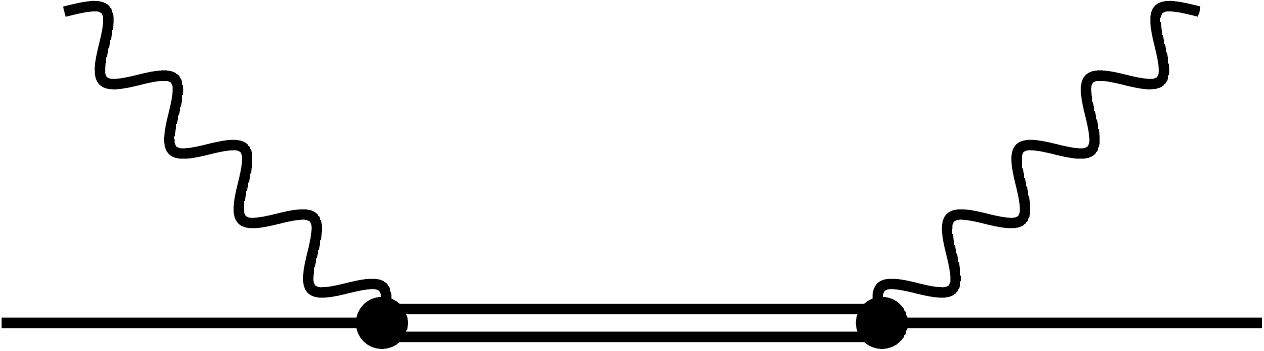}
 \caption{\small{Tree-level $\Delta(1232)$-exchange diagram. \label{DeltaExchange}}}
\end{figure}

We also include the Coulomb-quadrupole $(C2)$ $N\to \Delta$ transition,
described by the $g_C$ term in the following non-minimal $\gamma^* N \Delta$ coupling~\cite{Pascalutsa:2005vq,Pascalutsa:2005ts}
(note that in these references the overall sign of $g_C$ is inconsistent between the Lagrangian and Feynman rules):
\bea
\mathcal{L}^{(2)}_\Delta &=&  \frac{3e}{2M_N M_+}\,\overline N\, T_3\,\Big\{
i g_M  \tilde F^{\mu\nu} \,\partial_{\mu}\Delta_\nu- g_E \gamma_5 F^{\mu\nu}\,\partial_{\mu}\Delta_\nu\eqlab{gammaNDeltaLag}\\
&& +i \frac{g_C}{M_\Delta}\gamma_5 \gamma^\alpha (\partial_\alpha \Delta_\nu-\partial_\nu \Delta_\alpha)\partial_\mu F^{\mu \nu}\Big\}+\,\mbox{H.c.},
\nn
\eea
with $M_+=M_N+M_\Delta$ and the dual of the electromagnetic field strength tensor
 $\tilde F^{\mu\nu}=\frac{1}{2}\epsilon^{\mu\nu\rho\lambda}F_{\rho\lambda}$.
The electric, magnetic and Coulomb couplings ($g_E$, $g_M$ and $g_C$)
are known from the analysis of pion photoproduction $P_{33}$ multipoles~\cite{Pascalutsa:2005ts}.
The corresponding numerical values, as well as those of other physical constants used in this work, are given in Table~\ref{tab:constants}. The Coulomb coupling is subleading
compared with the electric and magnetic couplings, and it was not included in the previous calculations.
However, the relatively large magnitude of $g_C$ hints at its potential numerical importance,
which we examine in this work.

The counting of the $\De(1232)$ effects is done within  ``$\delta$-counting''~\cite{Pascalutsa:2003aa}, where the Delta-nucleon mass difference,
$\varDelta=M_\Delta-M$, is a light scale ($\varDelta \ll \La_\chi$) that is
substantially heavier than the pion mass ($m_\pi\ll\varDelta)$. Hence, if $p\sim m_\pi$, 
then $\mathcal{O}(p^4/\varDelta)$ is in between of $\mathcal{O}(p^3)$ and $\mathcal{O}(p^4)$.

For the non-Born VVCS amplitudes
and polarizabilities the predictive orders are $\mathcal{O}(p^3)$ and $\mathcal{O}(p^4/\varDelta)$.
The $\mathcal{O}(p^3)$ contribution comes from the pion-nucleon ($\pi N$) loops.
We refer to it here as the LO contribution.\footnote{In the full Compton amplitude, there is a lower order contribution coming from the Born terms, leading to a shift in nomenclature by one order: the LO contribution referred to as the NLO contribution, etc., see e.g.~Ref.~\cite{Lensky:2009uv}. }
The $\mathcal{O}(p^4/\varDelta)$ contribution, arising at the NLO, 
comes from the tree-level Delta-exchange ($\Delta$-exchange) graph shown in  Fig.~\ref{DeltaExchange}, and the pion-Delta ($\pi \Delta$) loops. The loop diagrams are shown in Ref.~\cite[Figs.~1 and 2]{Lensky:2014dda}.

The $\Delta$-exchange graph is described by the $\gamma^* N \Delta$ interaction in \Eqref{gammaNDeltaLag}. For the magnetic coupling, one assumes a dipole behavior to mimic the form expected from 
vector-meson dominance (VMD):
\begin{equation}
g_M\to \frac{g_M}{\big[1+Q^2/\Lambda^2\big]^2}\,,\eqlab{modifiedgm}
\end{equation}
with the dipole mass $\Lambda^2=0.71$~GeV${^2}$.
 This modification is going beyond the standard $\chi$PT framework, although it may in principle be implemented 
within $\chi$PT by systematic inclusion of vector mesons, as is done for the nucleon form factors in, e.g., Ref.~\cite{Schindler:2005ke}. Another possibility is to represent the above VMD effect by a 
resummation of the 
$\pi \pi$ rescattering diagrams in the $t$-channel. 
In either case, the inclusion of this $Q^2$ dependence is crucial for the correct description of the pion electroproduction data \cite{Pascalutsa:2005vq}. Since the pion electroproduction is, via the sum rules, affecting the polarizabilities, it can be expected that a better description of the electroproduction data leads to a better description of the $Q^2$ behavior of polarizabilities. The effect of this modification of $g_M$ is illustrated in Figs.~\ref{Fig:alpha+beta-orders}, \ref{Fig:M14-orders},
and \ref{fig:subtraction}.

\begin{table}[b]
\caption{Parameters (fundamental and low-energy constants) \cite{Agashe:2014kda} at the order they first appear. The $\pi N\Delta$ coupling constant $h_A$ is fit to the experimental Delta width and the $\gamma^* N \Delta$ coupling constants $g_M$, $g_E$ and $g_C$ are taken from the pion photoproduction study of Ref.~\cite{Pascalutsa:2005vq}. The free parameters $\de\be_{p,n}$ are fitted to the Baldin sum rule for the proton and neutron \cite{Gryniuk:2015aa,Levchuk:1999zy}, respectively.\label{tab:constants}} 
\begin{tabular}{lp{0.3cm}l}
\hline\\
$\mathcal{O}(p^2)$&&$\alpha\simeq 1/(137.04)$, $M_N=M_p\simeq 938.27$ MeV\\
$\mathcal{O}(p^3)$&&$g_A\simeq 1.27$, $f_\pi\simeq 92.21$ MeV, $m_\pi\simeq 139.57$ MeV \\
$\mathcal{O}(p^4/\varDelta)$&&$M_\Delta\simeq 1232$ MeV, $h_A\equiv 2g_{\pi N \Delta}\simeq 2.85$, $g_M\simeq 2.97$, $g_E\simeq -1.0$, $g_C\simeq -2.6$\\
$\mathcal{O}(p^4)$ && $\de\be_p = -1.12\times 10^{-4}\text{ fm}^3$, $\de\be_n = -3.10\times 10^{-4}\text{ fm}^3$\\
\\
\hline
\end{tabular}
\end{table}

A feature of the $\delta$-counting is that the characteristic momentum
$p$ distinguishes two regimes: the {\it low-energy} ($p\simeq m_\pi$) and 
{\it resonance} ($p\simeq \varDelta$) regimes. The above counting is limited to the 
low-energy regime.
Since we are interested in the low-energy expansion of the  VVCS amplitudes (i.e., the expansion in powers of small $\nu$ with $Q^2$ finite),
we do not consider the regime where one-Delta-reducible graphs are enhanced (resonance regime).
However, going to higher $Q^2$ one does need to count the 
Delta propagators similar to the nucleon propagators, which, in turn, calls for inclusion
of $\pi \Delta$ loops with two and three Delta propagators, which have been omitted here. They are only included implicitly 
 by adjusting the isospin coefficients of the one-nucleon-reducible $\pi \Delta$-loop graphs to restore current conservation, as explained in the next section. Apart from that, $\pi \Delta$ loops have a rather mild
dependence on momenta and the missing loops are unlikely to  affect the
$Q^2$-dependence of the moments of structure functions significantly, even for $Q^2$ comparable to $\varDelta^2$.

\subsection{Renormalization}

The calculation of the $\pi N$- and $\pi\Delta$-loop graphs is analogous to Ref.~\cite{Lensky:2009uv}, with the obvious extension to the case of a finite photon virtuality. The renormalization is also done in the exact same way; 
namely, 
subtracting the loop contribution to the Born term of the the VVCS amplitude.
The $\pi \Delta$-loop graphs
still contain divergences after this subtraction. These divergences are of higher orders, $\mathcal{O}(p^5/\varDelta^2)$ and  $\mathcal{O}(p^4)$,
and will be canceled by the corresponding higher-order
contact terms. In practice, they are removed by taking the $\overline{\text{MS}}$ values of the divergent quantities.

As mentioned above, $\pi \Delta$-loop graphs where photons couple minimally to the Delta contain more 
than one Delta propagator and therefore should be suppressed by extra powers of $p/\varDelta$.
However, their lower-order contributions are important for electromagnetic gauge invariance and therefore for the renormalization procedure. This issue is similar to HB$\chi$PT, where $\pi N$ loops with nucleon-photon couplings are suppressed (in the Coulomb gauge) and not included at $\mathcal{O}(p^3)$, even though they are required for manifest electromagnetic gauge invariance. It is then said that the gauge-invariance violating pieces are of higher order. Here, in the $\delta$-counting, we choose to retain exact gauge invariance, by means of including a minimal set of higher-order contributions. 

This is achieved, as first done in~\cite{Lensky:2009uv}, by observing that a gauge-invariant set of 
diagrams with one Delta propagator arises for the particular case of neutral Delta, 
$\De^0$. The ratio of the isospin factors between the one-particle-irreducible (1PI) and 
one-particle-reducible (1PR) graphs is then set to correspond with the $\De^0$ case.
This procedure ensures exact gauge invariance and the low-energy theorem~\cite{Low:1954kd, GellMann:1954kc}, thus facilitating the correct renormalization of the charge and anomalous magnetic moment of the nucleon.
In this way one
includes the relevant contributions of the omitted one-loop graphs with minimal coupling of photons to the Delta. When  the latter graphs 
are included explicitly in a future higher-order calculation, the isospin factors of 1PR graphs will be restored to actual values.

\subsection{Uncertainty estimate}

To estimate the uncertainties of our NLO predictions, we define the running expansion parameter
\begin{align}
 \tilde{\delta}(Q^2) = \sqrt{ \left(\frac{\varDelta}{M_N}\right)^2 + \left(\frac{Q^2}{2 M_N \varDelta}\right)^2 },\eqlab{dtilde}
\end{align}
such that the next-to-next-to-leading order (N$^2$LO) is expected to be of relative size $\tilde{\delta}^2$ \cite{Pascalutsa:2005vq}.
To estimate the uncertainty of a polarizability $P(Q^2)$ due to the neglected higher-order terms in the
chiral expansion, we separate that polarizability into the real-photon piece
$P(0)$ and the $Q^2$-dependent remainder $P(Q^2)-P(0)$. The uncertainty of $P(Q^2)$ is obtained by adding the
estimates for these two parts in quadrature:
\beq
\Delta P(Q^2)= \sqrt{\tilde \delta^4(0) P(0)^2 +\tilde \delta^4(Q^2) \left[P(Q^2)-P(0)\right]^2},
\eeq
The uncertainties in the values of the parameters have a much smaller impact compared to the truncation
uncertainty and are therefore neglected.

\section{Results and discussion}

We now consider the numerical results for some of 
the moments of the nucleon structure functions that appear in the expansion \Eqref{VVCSexpansion}. We shall also consider the proton subtraction function $\ol{T}_1(0,Q^2)$. The complete NLO values will be decomposed into three
individual contributions: the $\pi N$ loops, the $\Delta$ exchange, and the $\pi \Delta$ loops. In practice, we extract all results from the calculated non-Born VVCS amplitudes. For a cross-check, we used the photoabsorption cross sections  described in App.~\ref{App:CrossSections}.

\label{Sec:Scalar-Pol}

\subsection{\boldmath{$M_1^{(2)}(Q^2)$} --- the generalized Baldin sum rule}
\label{M12genBS}

\begin{figure}[hbt]
\begin{center}
 \includegraphics[width=0.49\textwidth]{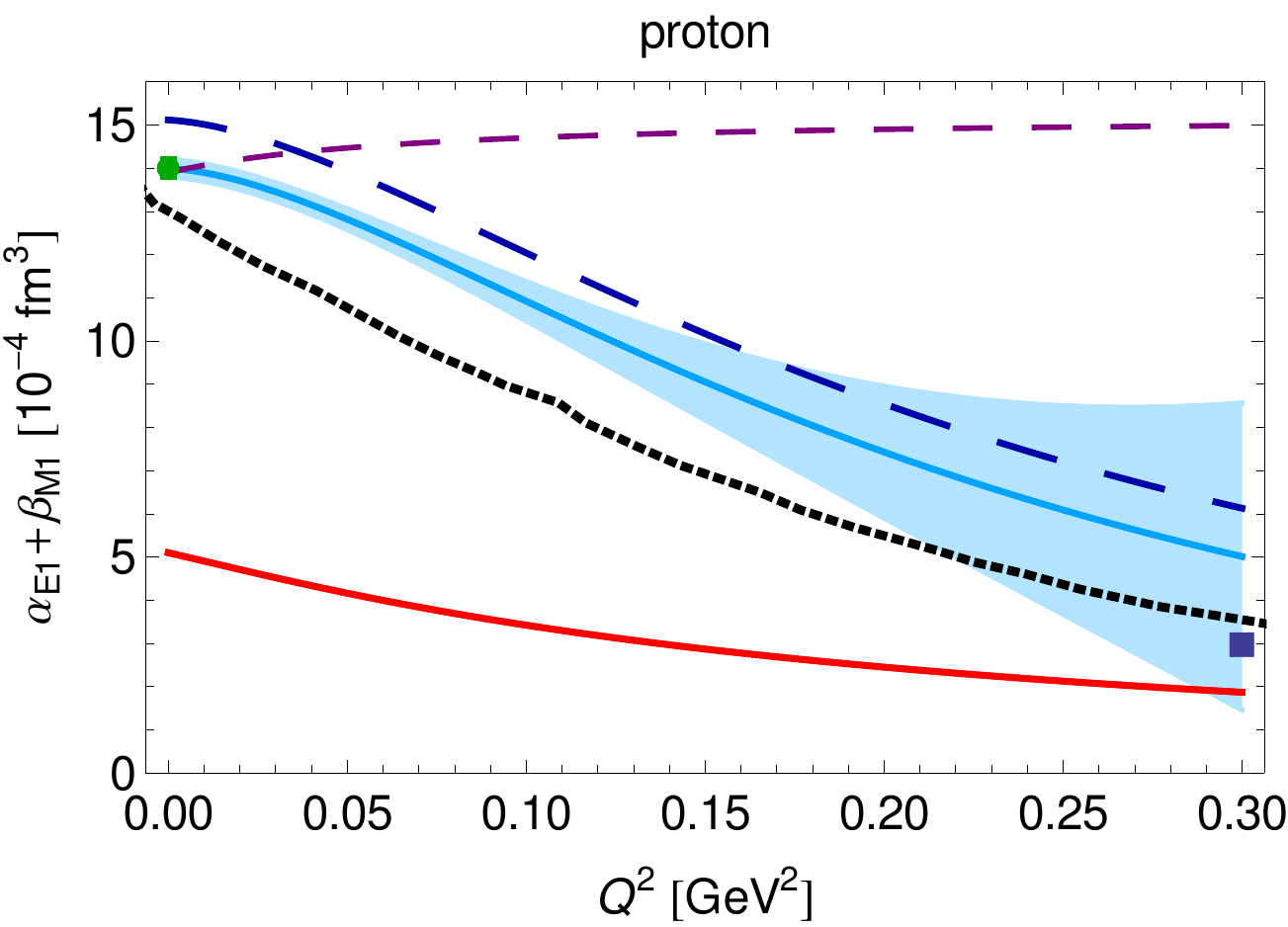}
  \includegraphics[width=0.49\textwidth]{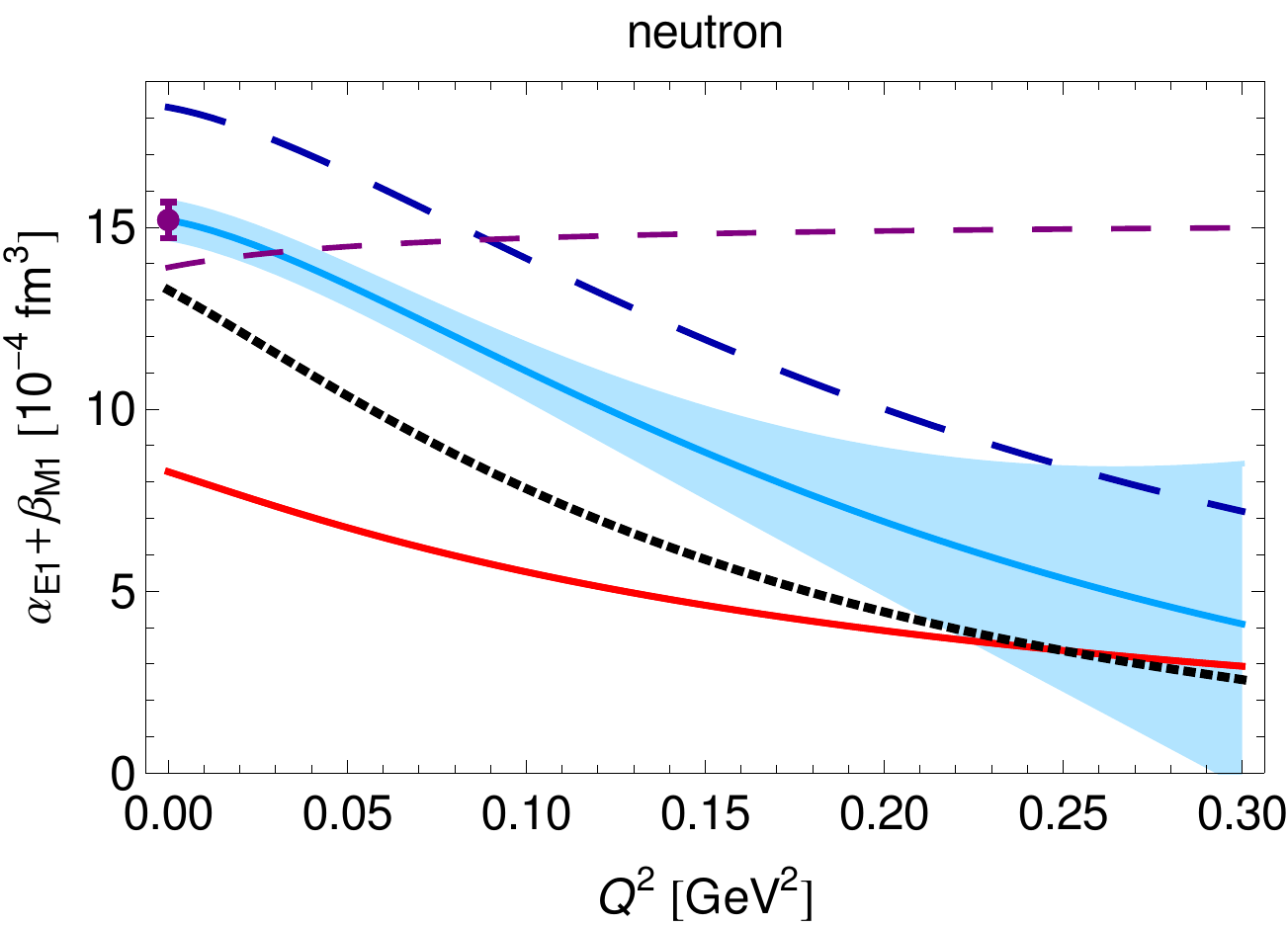}\\[0.2cm]
   \includegraphics[width=0.49\textwidth]{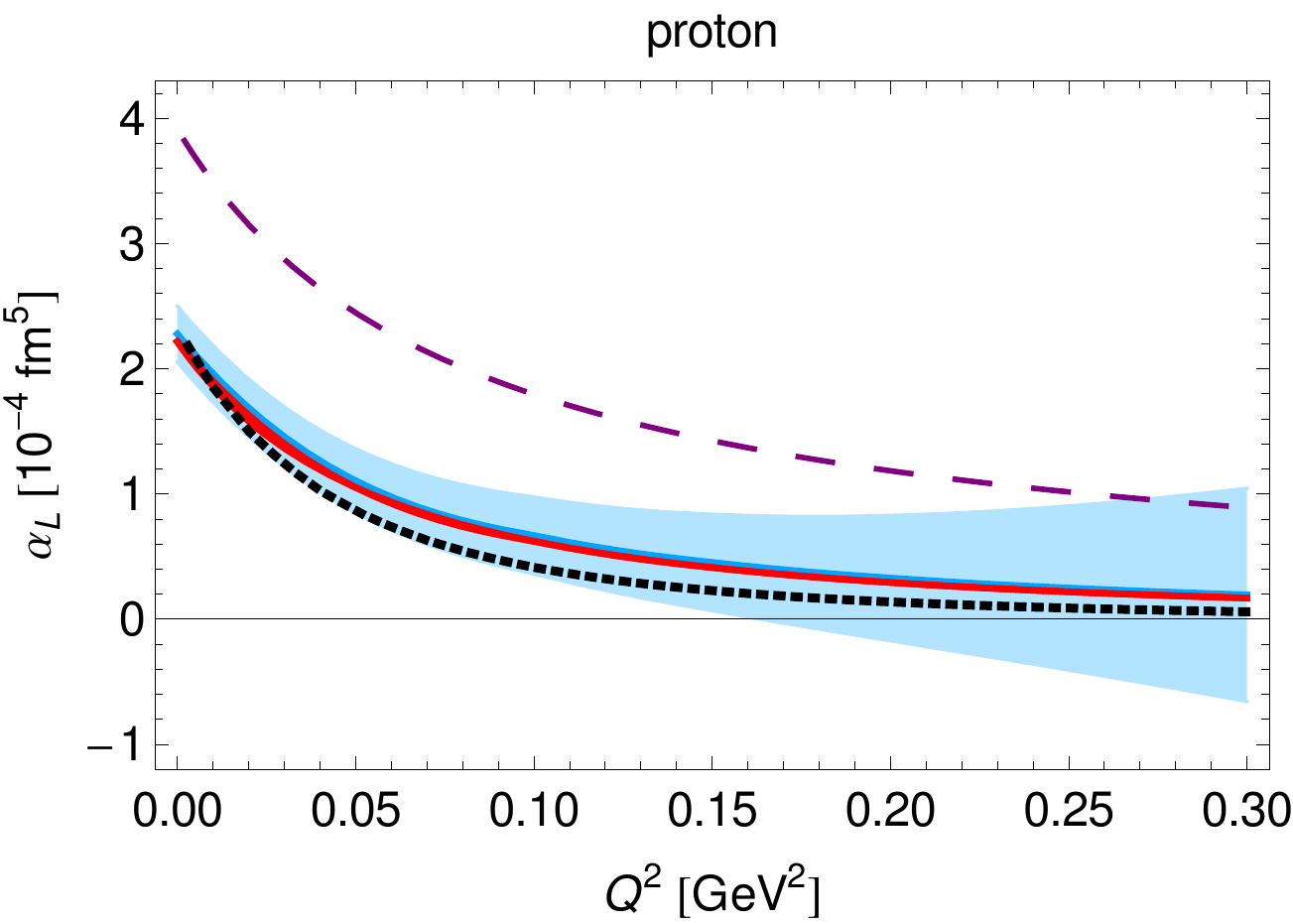}
    \includegraphics[width=0.49\textwidth]{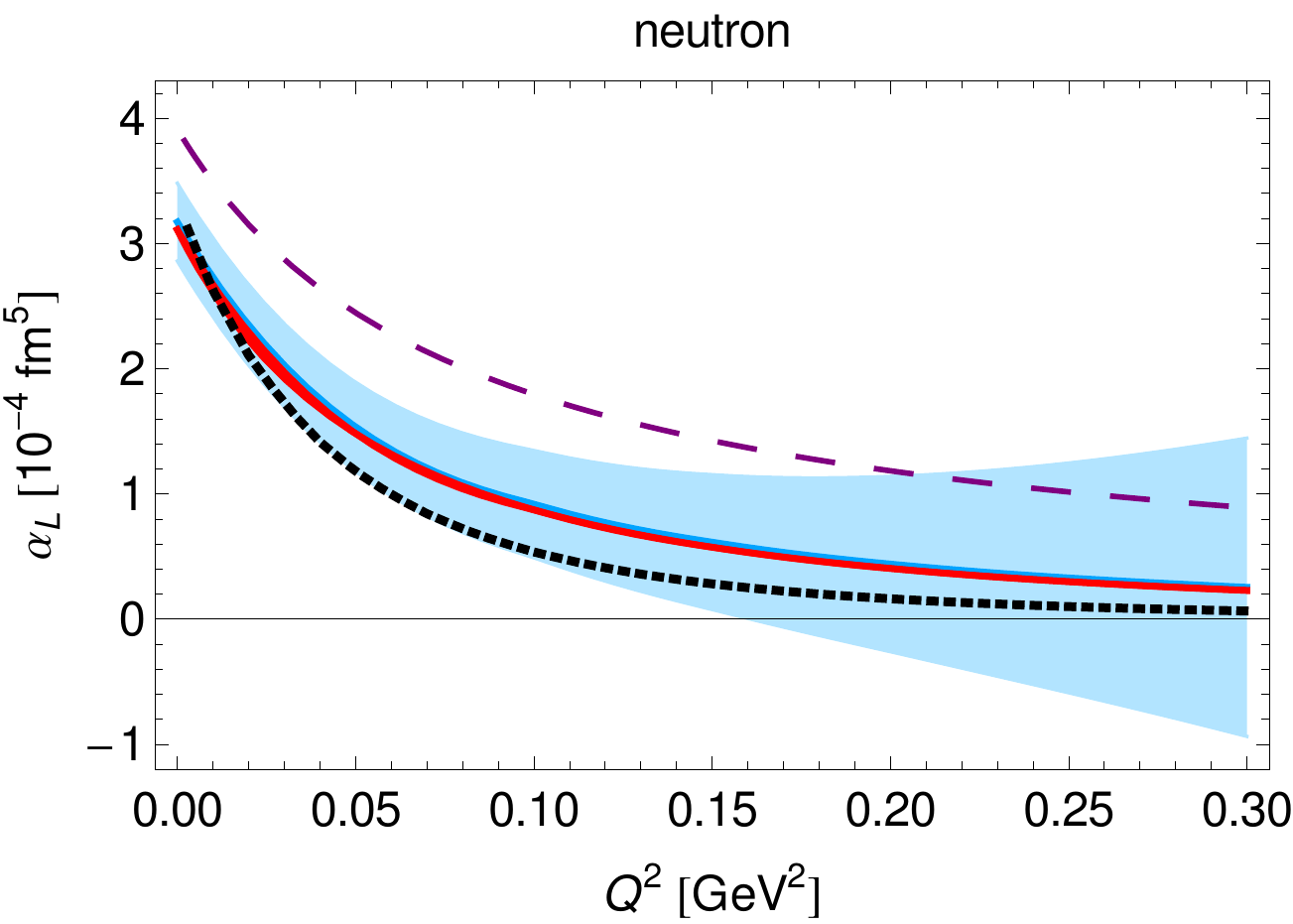}
\caption{\small{Upper panel: Generalized Baldin sum rule for the proton (left) and neutron (right) as function of $Q^2$.  
The result of this work, including the $\delta\beta$ contribution, is shown by the blue solid line, with the blue band representing the uncertainty due to higher-order effects. The blue long-dashed line shows the NLO B$\chi$PT prediction (i.e., without the $\delta\beta$ term).
The red line represents the LO B$\chi$PT result, while the purple dashed line is the $\mathcal{O}(p^3)$ HB result~\cite{Nevado:2007dd}.
The black dotted line is the MAID model prediction~\cite{Drechsel:2000ct,Drechsel:1998hk,private-Lothar};  for the proton we use the updated estimate from Ref.~\cite{Drechsel:2002ar} that includes the $\pi, \eta, \pi\pi$ channels.
At $Q^2=0$~GeV$^2$, we show the Baldin sum rule value for the proton (green dot)~\cite{Gryniuk:2015aa} and neutron (purple dot)~\cite{Levchuk:1999zy}.
 For the proton, the $Q^2=0.3$~GeV$^2$ point (blue square) is the empirical evaluation of Ref.~\cite{Liang:2004tk}. Lower panel: Longitudinal polarizability for the proton (left) and neutron (right). The NLO B$\chi$PT predictions of this work are shown by the blue solid line with blue band; the legend for the remaining curves is as in the upper panel. Note that the LO B$\chi$PT curves are practically on top of the NLO ones.}\label{Fig:alpha+betaplot}}
\end{center}
\end{figure}

The electric and magnetic dipole polarizabilities, $\alpha_{E1}(Q^2)$ and $\beta_{M1}(Q^2)$, encode information about the dipole response of the nucleon to an electromagnetic field. For finite momentum transfers, the sum of dipole polarizabilities is given by the generalized Baldin sum rule:
\beq
[\alpha_{E1}+\beta_{M1}] (Q^2)= \frac{1}{2 \pi^2} \int_{\nu_0}^\infty \! \dd\nu\,\sqrt{1+\frac{Q^2}{\nu^2}}\, \frac{\sigma_T (\nu,Q^2)}{\nu^2} =\frac{8 \al M_N}{Q^4}\int_0^{x_0}\!\dd x\, x \,F_1(x,Q^2),\label{Eq:alpha+betaQ2}
\eeq
where $\nu_0$ is the lowest inelastic threshold, in this case the one-pion production threshold $\nu_0=m_\pi + (m_\pi^2+Q^2)/2M_N$, and $x_0=Q^2/2M_N \nu_0$.
The electric and magnetic dipole polarizabilities of the nucleon enter the nucleon-structure contributions
to the Lamb shift of muonic hydrogen and other muonic atoms \cite{Bernabeu:1982qy,Pachucki:1996zza,Carlson:2011zd,Alarcon:2013cba}, and thus
are of major interest for an accurate extraction of the nuclear charge radii.

Our B$\chi$PT predictions for $\alpha_{E1}+\beta_{M1}$ are shown in Fig.~\ref{Fig:alpha+betaplot} \{upper panel\}, for both the proton
and the neutron, up to photon virtualities of $0.3$~GeV$^2$. Our main result is given by the blue solid lines and the error bands, where we used the $p^4$ LEC $\delta\beta$ to fit the static polarizabilities to the empirical Baldin sum rule values (green and purple dots) given in \Eqref{fittedBaldin}, see discussion in Sec.~\ref{sec:PowerCounting}. The inclusion of
$\delta\beta$, cf.\ Table \ref{tab:constants}, merely leads to a constant shift, as can be seen by comparing to the pure $\mathcal{O}(p^4/\Delta)$ predictions (blue long-dashed lines), which include the $\pi N$-loop, the $\Delta$-exchange and the $\pi \Delta$-loop contributions.  To illustrate the effect of the Delta in these predictions, we also plot the LO $\pi N$-loop contributions
separately (red solid lines). We compare our results for the $Q^2$
evolution with the $\mathcal{O}(p^3)$ HB$\chi$PT predictions~\cite{Nevado:2007dd} (purple dashed lines) and the MAID model predictions~\cite{Drechsel:2000ct,Drechsel:1998hk} (black dotted lines). The latter are based on the generalized Baldin sum rule~\eqref{Eq:alpha+betaQ2} evaluated with ($\pi+\eta+\pi \pi$) photoproduction
cross sections~\cite{Drechsel:2002ar}. 
The data points are also evaluations of the (generalized) Baldin sum rule \cite{Liang:2004tk,Gryniuk:2015aa,Babusci:1997ij}. One can see that the B$\chi$PT predictions seem to systematically overestimate the MAID model in the $Q^2$ range shown here. One has to note that
the MAID model, on the other hand, slightly underestimates the empirical Baldin sum rule evaluations.

The $\mathcal{O}(p^3)$ HB results seem to agree with the empirical values at the real-photon point~\cite{Babusci:1997ij}
 for both the proton and the neutron. However, they do not fall off with increasing $Q^2$ in contrast to the B$\chi$PT predictions.
This asymptotic behavior is the reason for the large proton-polarizability effect on the muonic-hydrogen Lamb shift found within HB$\chi$PT \cite{Nevado:2007dd,Peset:2014yha,Peset:2014jxa}, much larger than the phenomenological value. As shown in Refs.~\cite{Alarcon:2013cba,Lensky:2017bwi}, this issue is solved within the relativistic formulation, which gives a result closer to calculations based on the dispersive approach.

\begin{figure}
\begin{center}
 \includegraphics[width=0.49\textwidth]{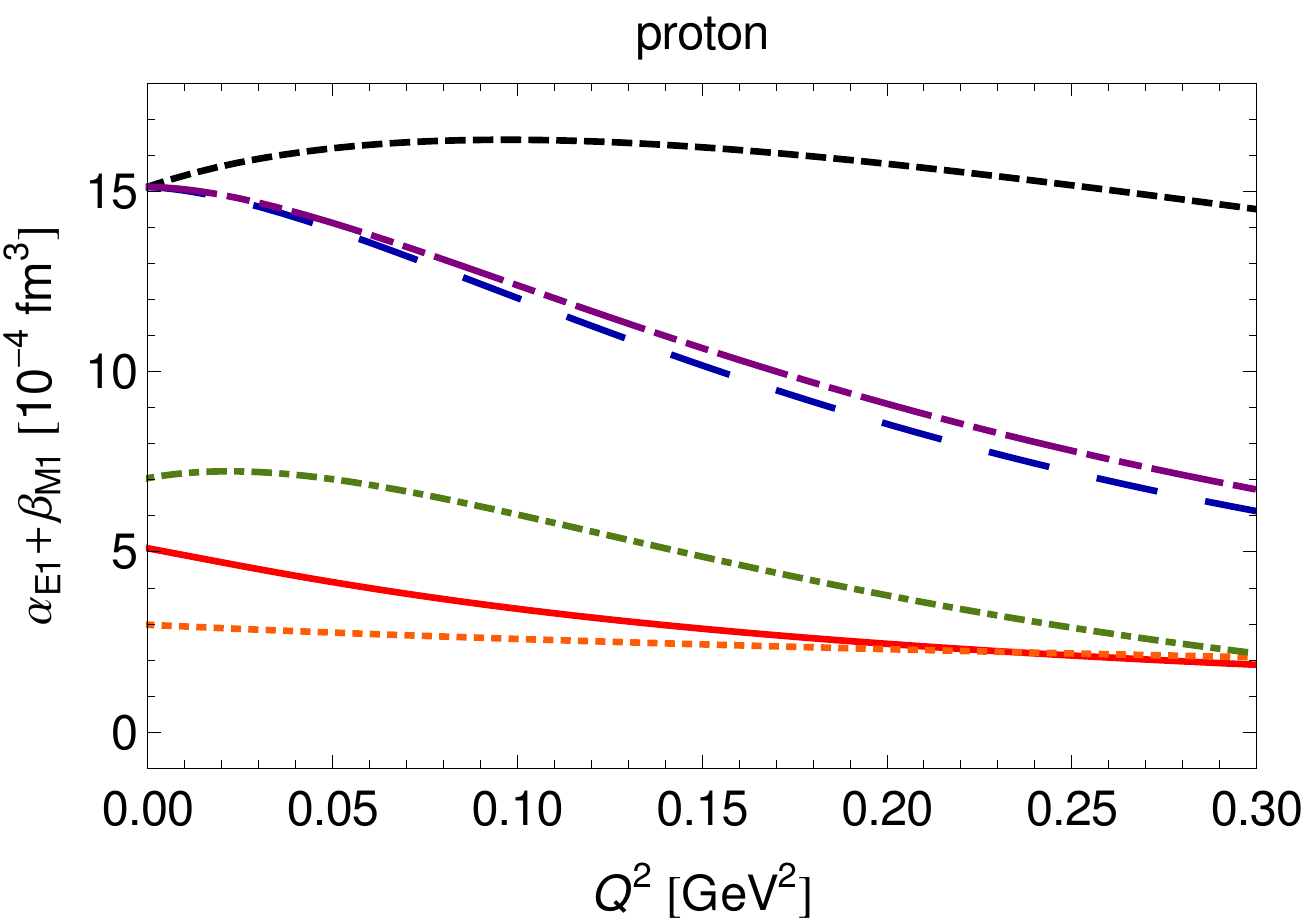}
  \includegraphics[width=0.49\textwidth]{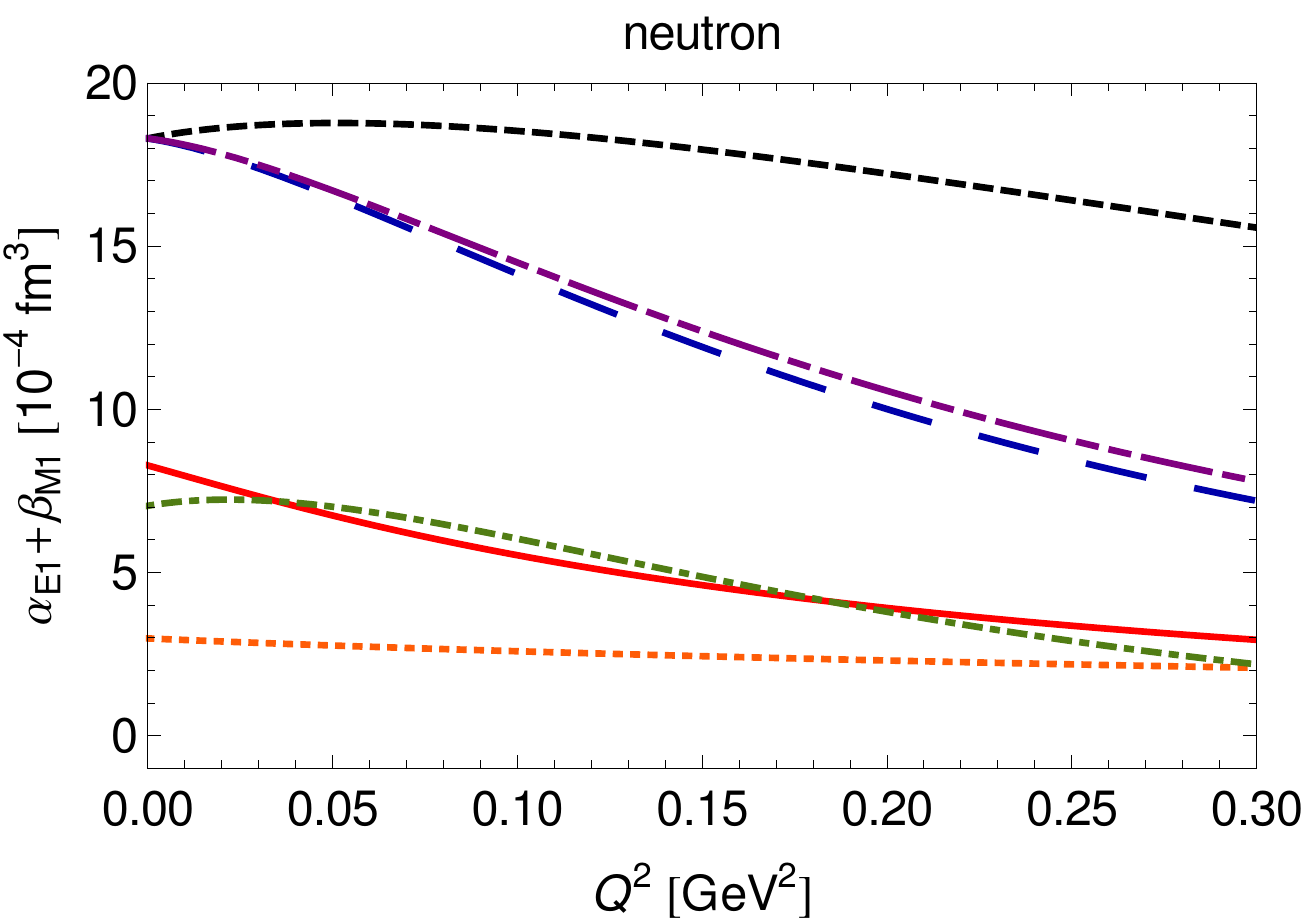}\\[0.2cm]
   \includegraphics[width=0.49\textwidth]{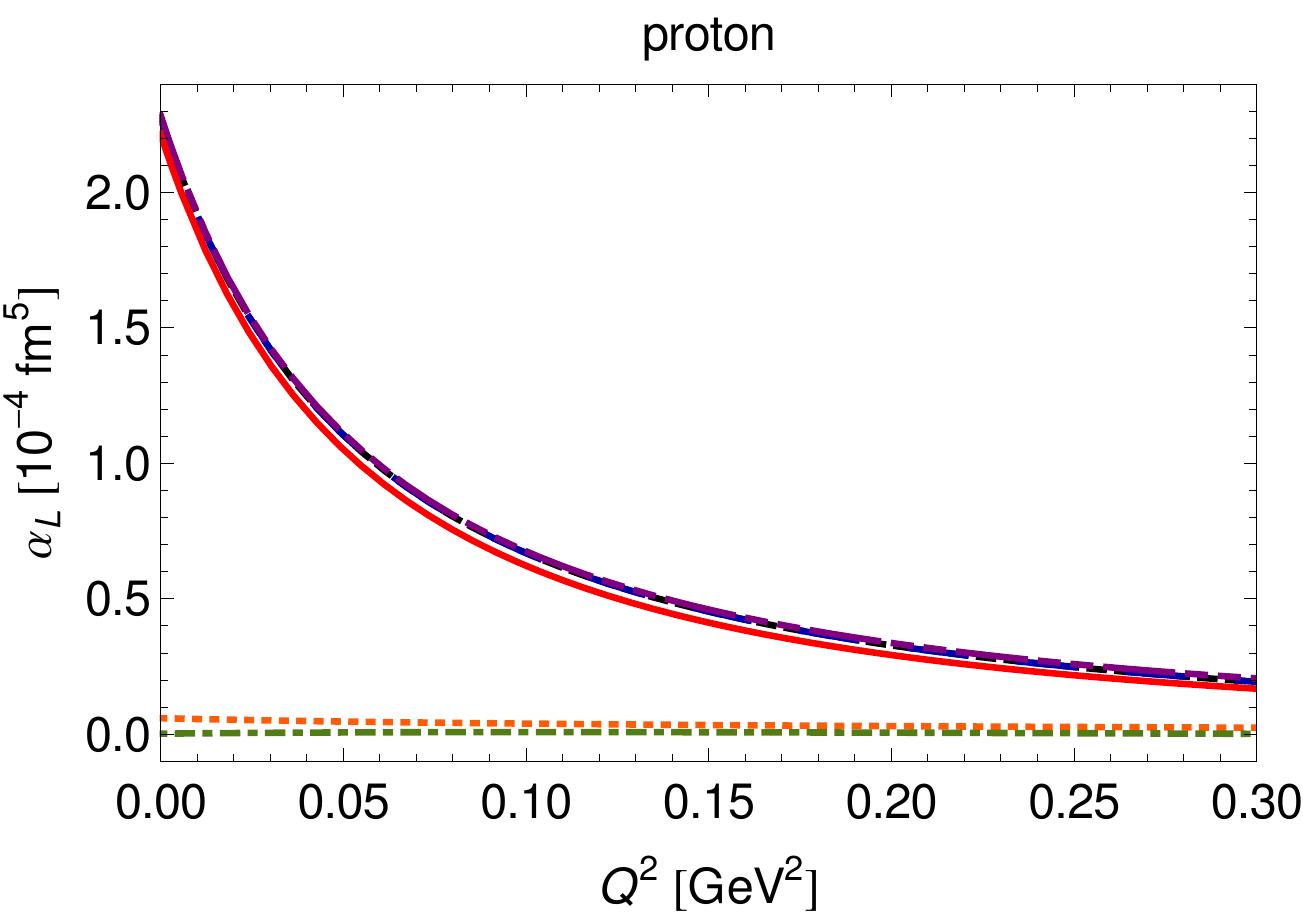}
    \includegraphics[width=0.49\textwidth]{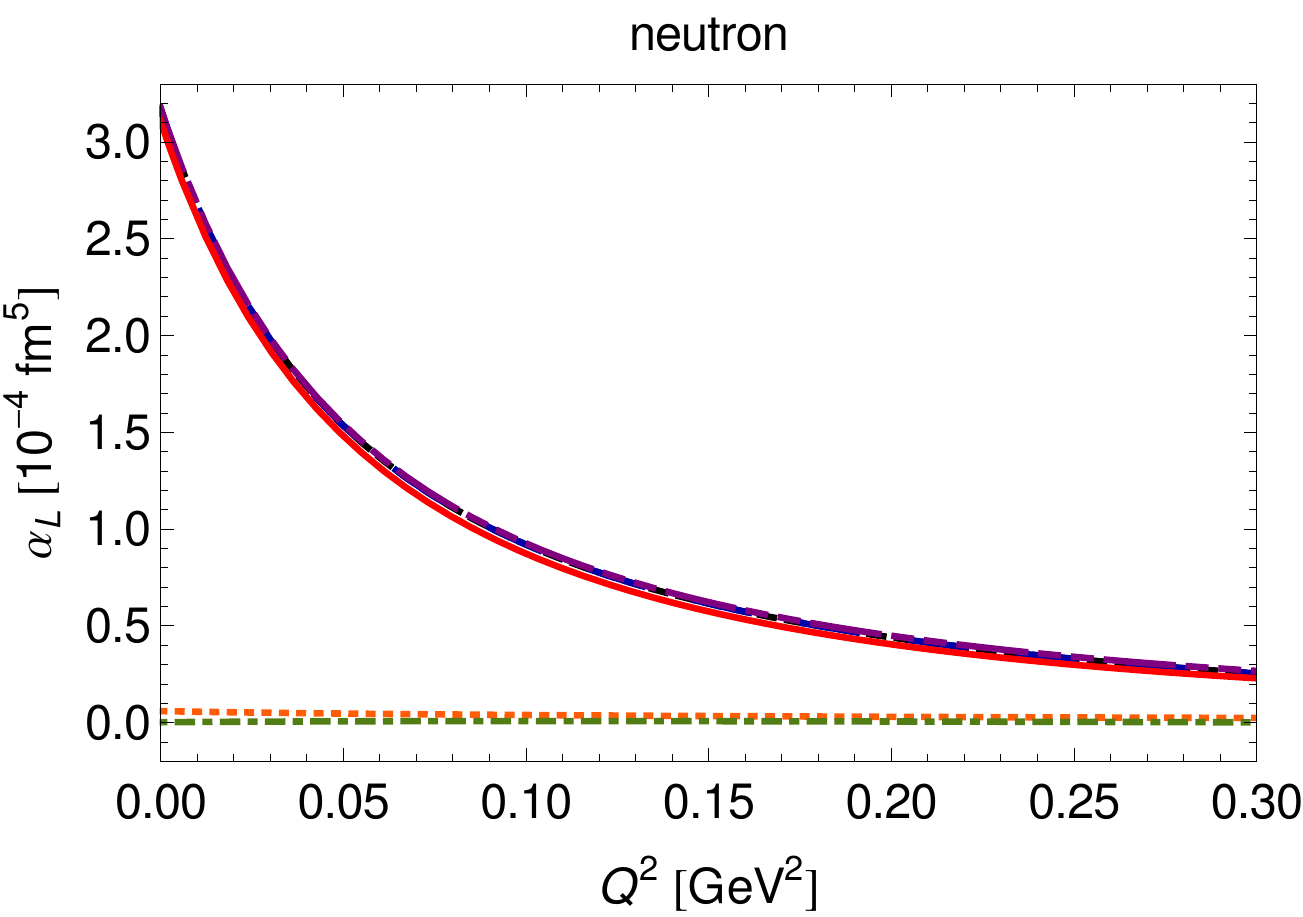}
\caption{\small{Contributions of the different orders to the chiral predictions of $[\alpha_{E1}+\beta_{M1}](Q^2)$ \{upper panel\} and $\alpha_L(Q^2)$ \{lower panel\} for the proton (left) and neutron (right). Red solid line: $\pi N$-loop contribution, green dot-dashed line: $\Delta$-exchange contribution, orange dotted line: $\pi \Delta$-loop contribution, blue long-dashed line: total result, purple dot-dot-dashed line: total result without $g_C$ contribution, black short-dashed line: total result without $g_M$ dipole.}}\label{Fig:alpha+beta-orders}
\end{center}
\end{figure}

The static dipole polarizabilities $\alpha_{E1}$ and $\beta_{M1}$ have been  studied  within both the HB  and
the B$\chi$PT. While  HB$\chi$PT gives results
remarkably close to the experimental determinations already at LO \cite{Bernard:1995dp},
the contribution of the $\Delta(1232)$ is harder to accommodate in this framework~\cite{Hemmert:1996rw}. In contrast to that, LO  B$\chi$PT \cite{Bernard:1991rq,Bernard:1991ru}
yields smaller values for the sum of dipole polarizabilities, in disagreement with the empirically extracted values based on evaluations of the Baldin sum rule with modern photoabsorption data~\cite{Babusci:1997ij,Olm01,Gryniuk:2015aa}. However, the NLO contributions from $\Delta$ exchange and $\pi\Delta$ loops improve the situation \cite{Lensky:2009uv,Lensky:2015awa}.
In the case of the proton, they bring the B$\chi$PT result in agreement with the experimental extraction, while for the neutron the total result is slightly bigger. The $\Delta(1232)$ contributions are, therefore, naturally accommodated in B$\chi$PT, and not in HB$\chi$PT (where they can be reconciled with the empirical values only by means of the $p^4$ LEC $\delta\beta$, see, e.g., Refs. \cite{Griesshammer:2012we,McGovern:2012ew} for the recent calculations and review).

The B$\chi$PT contributions from $\pi N$ loops, $\Delta$ exchange and $\pi \Delta$ loops to the dipole polarizabilities are, in that order and in the usual units of $10^{-4}$~fm$^3$:
\begin{subequations}
\label{Eq:alpha+betaRealPoint}
\begin{align}
\alpha_{E1p}+\beta_{M1p}=15.12(1.48) \approx 5.10+7.04+2.98, \label{Eq:alpha+betaProtonRealPoint}\\
\alpha_{E1n}+\beta_{M1n} =18.30(1.79)\approx 8.28+7.04+2.98. \label{Eq:alpha+betaNeutronRealPoint}
\end{align}
\end{subequations}
This NLO result is a prediction of B$\chi$PT, i.e., it does not include the $p^4$ LEC $\delta\beta$ discussed in Sec.~\ref{sec:PowerCounting}.
The corresponding individual contributions to the $Q^2$-dependent generalized polarizabilities are shown in Fig.~\ref{Fig:alpha+beta-orders} \{upper panel\}.
For the proton, the dominant contribution in the studied $Q^2$ range is that of the $\Delta$ exchange, while for the neutron the $\pi N$-loop and $\Delta$-exchange contributions are of roughly the same size.
The importance of the Delta is related to the fact that the nucleon-to-Delta transition is dominantly of the magnetic dipole type, and therefore it gives a huge contribution to $\beta_{M1}$.

In addition, we investigate the slopes of the polarizabilities at the real-photon point. Decomposing the results as before into the three contributions, we observe that B$\chi$PT predicts large contributions to the slopes both from $\pi N$ loops and $\Delta$ exchange. The $Q^2$ dependence generated by $\pi \Delta$ loops, on the other hand, is negligible,
as can  clearly be seen from Fig.~\ref{Fig:alpha+beta-orders}. The numerical values for the individual contributions to the slopes are, in units of $10^{-4}$~fm$^5$:
\begin{subequations}
\bea
\left.\frac{\dd(\alpha_{E1p} + \beta_{M1p}) (Q^2)}{\dd Q^2}\right|_{Q^2=0}&=&-0.19(6)\approx -0.74  + 0.74-0.20 ,\\
\left.\frac{\dd(\alpha_{E1n} + \beta_{M1n}) (Q^2)}{\dd Q^2}\right|_{Q^2=0}&=& -0.68(21)\approx-1.22  +0.74-0.20.
\eea
\end{subequations}
 The dipole form factor in the magnetic coupling $g_M$ generates the $Q^2$ falloff of the dipole polarizabilities, cf.\ Fig.~\ref{Fig:alpha+beta-orders}, which is also observed
in parametrizations of experimental cross sections~\cite{Hall:2014lea}. 
Because of cancellations between the $\pi N$-loop and the $\Delta$-exchange contributions,
 the dipole also crucially affects the overall sign of the slope, as can be seen in Fig.~\ref{Fig:alpha+beta-orders}. Note that due to these cancellations we estimate the relative error
 of the slope by $\tilde{\delta}$ instead of $\tilde{\delta}^2$.

Evaluating the Baldin sum rule radius, 
\begin{align}\label{Eq:r2alphabetaDef}
r_{(\alpha+\beta)}^2\equiv - \frac{6}{\alpha_{E1}+\beta_{M1}}\left.\frac{\dd}{\dd Q^2}[\alpha_{E1}+\beta_{M1}](Q^2)\right|_{Q^2=0},
\end{align}
we obtain $r_{(\alpha+\beta)p}= 0.29(9)$~fm and $r_{(\alpha+\beta)n}=0.52(16)$~fm, 
where we estimated the relative error to be $\tilde{\delta}$. Here, we again used our result including the $\delta \beta$ contribution, i.e., we fixed the static polarizabilities to the Baldin sum-rule values in \Eqref{fittedBaldin}, while the slope is still a prediction of B$\chi$PT. 

The result for the proton is in tension with the sum-rule evaluations \cite{Liang:2004tk,Sibirtsev:2013cga,Hall:2014lea}, which use  empirical parametrizations of the structure function $F_1(x,Q^2)$, e.g.~\cite{Hall:2014lea}: 
\beq r_{(\alpha+\beta)p}=0.98(5)  \; \mbox{fm}.
\eeq 
From Fig.~\ref{Fig:alpha+betaplot} one can see that the MAID empirical parametrization also leads to a steeper slope than B$\chi$PT.
This calls for a careful revision of the low-momentum behavior of the empirical
parametrizations in the near future.

\subsection{\boldmath{$\alpha_L (Q^2)$} --- the longitudinal polarizability}

The low-energy expansion of the longitudinal VVCS amplitude goes as
\beq 
T_L(\nu, Q^2)/4\pi = \al_{E1} Q^2 + \al_L Q^2\nu^2 +\ldots
\eeq 
with $\al_L$ called the longitudinal polarizability. Note that, in terms
of the moments  $\alpha_L = M^{(1)\prime}_2(0)  - M^{(2)\prime}_1(0) 
+ M^{(4)}_1(0)$. The generalized longitudinal polarizability is given by,
\beq
\alpha_L (Q^2)= \frac{1}{2 \pi^2} \int_{\nu_0}^\infty\! \dd\nu\,\sqrt{1+\frac{Q^2}{\nu^{2}}}\,            \frac{\sigma_L (\nu,Q^2)}{Q^2\, \nu^{2}} = \frac{4 \al M_N}{Q^6}\int_0^{x_0}\! \dd x \, F_L(x,Q^2),\label{Eq:alphaLQ2}
\eeq
with
\beq
F_L(x,Q^2)=-2xF_1(x,Q^2)+\left(1+\frac{4M_N^2 x^2}{Q^2}\right)F_2(x,Q^2).\nn
\eeq
Our B$\chi$PT prediction for $\alpha_L(Q^2)$ is shown in Fig.~\ref{Fig:alpha+betaplot} \{lower panel\}, 
where we compare our results, with and without the Delta contributions, with the MAID model predictions~\cite{Drechsel:2000ct,Drechsel:1998hk,Drechsel:2002ar,private-Lothar} and the HB limit of the $\pi N$-loop contribution. One can see that the Delta plays a negligible role in the low-$Q^2$ evolution of $\alpha_L$, which in the B$\chi$PT approach is dominated by $\pi N$ loops.
Our results run very close to the MAID curves, with small discrepancies in the intermediate $Q^2$ region.
At higher virtualities, these discrepancies decrease.
The HB approach, on the other hand, seems to systematically overestimate the value of $\alpha_L$ in the considered $Q^2$ range. 
This relatively big mismatch can be traced back to the slow convergence of the $1/M_N$ expansion, as one can see from the analytic expression for the $\pi N$-loop contribution to $\alpha_L(Q^2=0)$ given in  App.~\ref{App:PolarizabilitiesAll}. 

For $\alpha_L$,
we obtain the following contributions from $\pi N$ loops,  $\Delta$ exchange and $\pi\Delta$ loops, in units of $10^{-4}$~fm$^5$:
\begin{subequations}
\begin{align}
\alpha_{Lp}=2.28(22)\approx 2.22+ 0.00+  0.06, \\
\alpha_{Ln}= 3.17(31)\approx 3.11 + 0.00+ 0.06.
\end{align}
\end{subequations}
For the slope at $Q^2=0$, we find, in units of $10^{-4}$~fm$^7$:
\begin{subequations}
\begin{align}
\left.\frac{\dd\alpha_{Lp} (Q^2)}{\dd Q^2}\right|_{Q^2=0}&=-1.63(16)\approx  -1.62 + 0.01 - 0.01  ,  \\
\left.\frac{\dd\alpha_{Ln} (Q^2)}{\dd Q^2}\right|_{Q^2=0}&= -2.25(22)\approx -2.24 + 0.01 - 0.01.
\end{align}
\end{subequations}
 The corresponding individual contributions to the $Q^2$ dependence
of $\alpha_L(Q^2)$ are demonstrated in Fig.~\ref{Fig:alpha+beta-orders} \{lower panel\}. One again notices
that  $\Delta$ exchange and  $\pi \Delta$ loops give negligible contributions in this $Q^2$ range. 
The smallness of the $\Delta$-exchange contribution is explained by the fact that the magnetic coupling $g_M$ does not contribute to $\al_L$.

\subsection{\boldmath{$M^{(1)}_2(Q^2)$} --- the first moment of $F_2(x,Q^2)$}

\begin{figure}[tbh]
\begin{center}
 \includegraphics[width=0.49\textwidth]{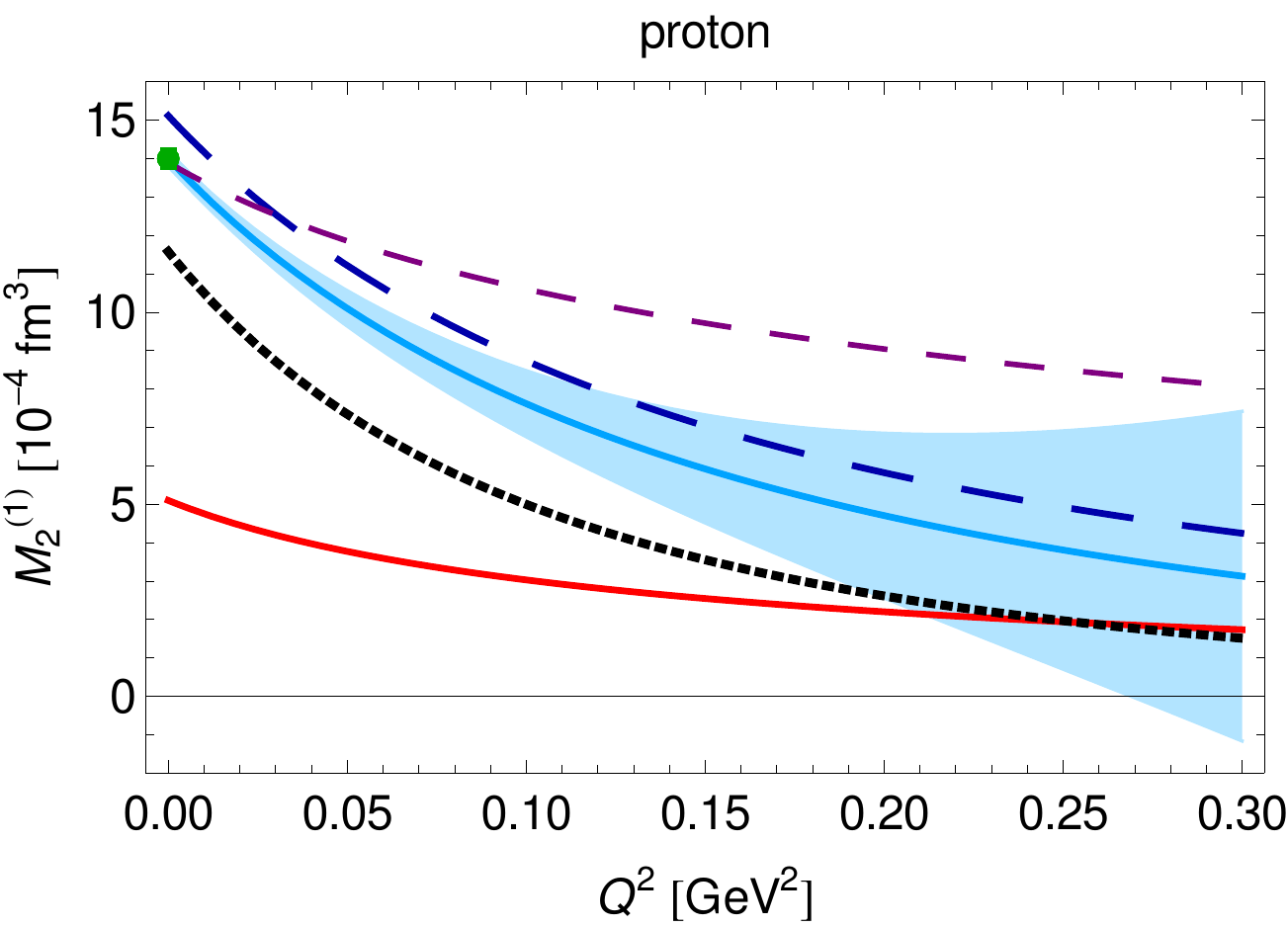}
  \includegraphics[width=0.49\textwidth]{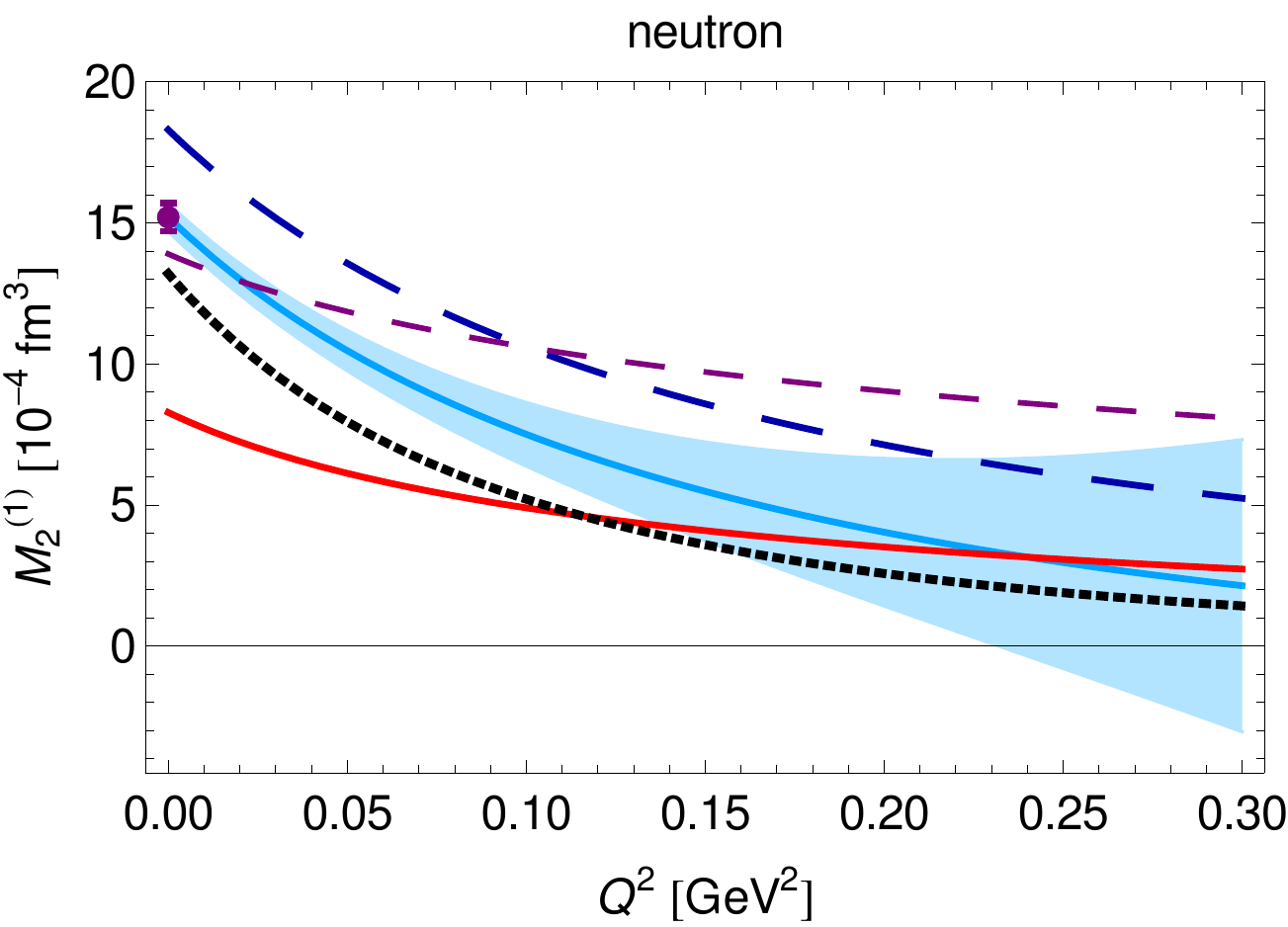}\\[0.2cm]
   \includegraphics[width=0.49\textwidth]{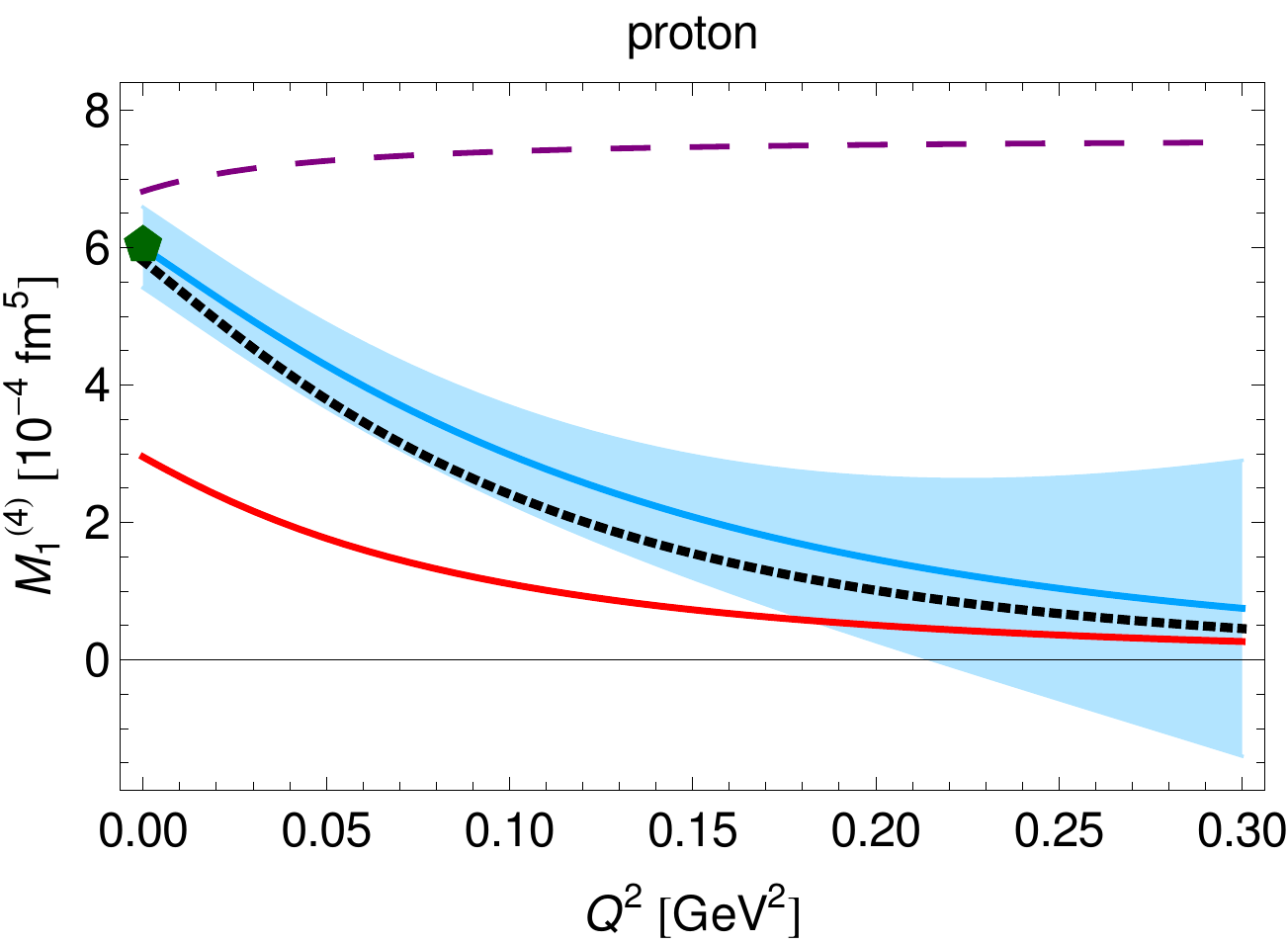}
    \includegraphics[width=0.49\textwidth]{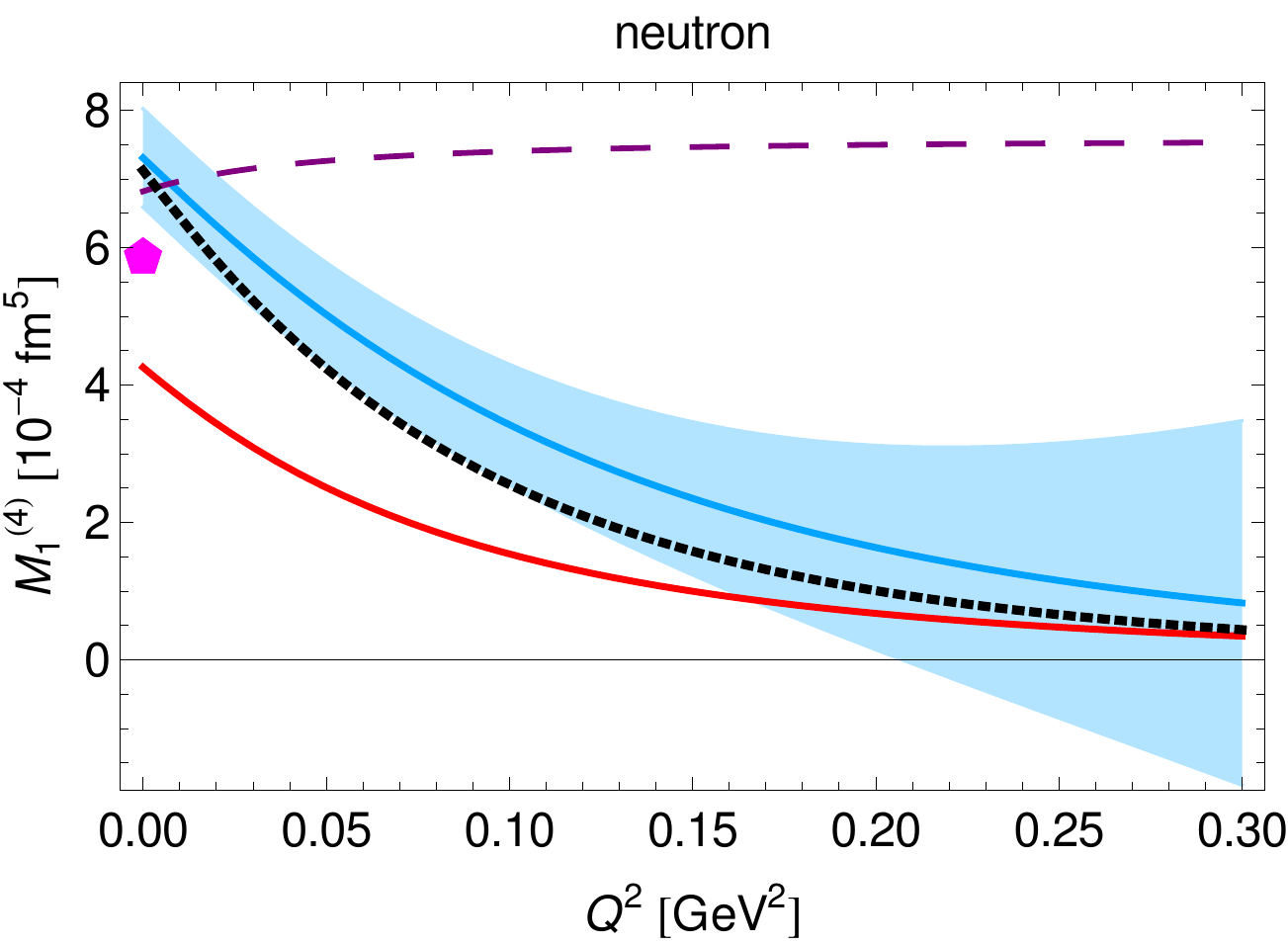}
\caption{\small{
Upper panel: The first moment of the structure function $F_2(x,Q^2)$ for the proton (left) and neutron (right) as function of $Q^2$.  
The result of this work, including the $\delta\beta$ contribution, is shown by the blue solid line, with the blue band representing the uncertainty due to higher-order effects. The blue long-dashed line shows the NLO B$\chi$PT prediction (i.e., without the $\delta\beta$ term).
The red line represents the LO B$\chi$PT result, while the purple dashed line is the $\mathcal{O}(p^3)$ HB result~\cite{Nevado:2007dd}.
The black dotted line is the MAID model prediction~\cite{Drechsel:2000ct,Drechsel:1998hk,private-Lothar}.
At $Q^2=0$~GeV$^2$, we show Baldin sum rule evaluations for the proton (green dot)~\cite{Gryniuk:2015aa} and neutron (purple dot)~\cite{Levchuk:1999zy}. Lower panel: Fourth-order generalized Baldin sum rule
for the proton (left) and neutron (right).
The NLO B$\chi$PT prediction of this work is shown by the blue solid line and the blue band;
the legend for the remaining curves is as in the upper panel.
At $Q^2=0$~GeV$^2$, we show fourth-order Baldin sum rule evaluations for the proton (dark green pentagon)~\cite{Gryniuk:2015aa} and neutron (magenta pentagon)~\cite{Schroder:1977sn}.}
 \label{Fig:M14plot}}
\end{center}
\end{figure}

At $Q^2=0$, the first moment of the structure function $F_2(x,Q^2)$,
\bea
M^{(1)}_2(Q^2)  &=& \frac{1}{2 \pi^2}  \int_{\nu_0}^{\infty}\, 
\frac{\mathrm{d}\nu}{\nu}  
\frac{1}{\sqrt{\nu^{2}+Q^2}}\left[ \frac{}{} \sigma_T(\nu, Q^2 ) +  \sigma_L(\nu, Q^2 )\right],   
\label{eq:m21}\\
&=&\frac{4 \al M_N}{ Q^4}\,\int_{0}^{x_0} \mathrm{d}x  \,F_2(x,\,Q^2), 
\nn 
\eea 
reproduces the Baldin sum rule:
$M^{(1)}_2(0) = \alpha_{E1}+\beta_{M1} = M^{(2)}_1(0)$, cf.\ Eq.~(\ref{Eq:alpha+betaRealPoint}). However, at finite
$Q^2$ this moment is independent of $M^{(2)}_1(Q^2)$. Comparing Fig.~\ref{Fig:M14plot} \{upper panel\},
which shows the $Q^2$ dependence of $M^{(1)}_2$, with the respective figure for $M_1^{(2)}$, Fig.~\ref{Fig:alpha+betaplot} \{upper panel\},
one can indeed see that the two moments noticeably diverge as one departs from the real-photon limit.
Note that the contribution of the $p^4$ operator in \Eqref{eq:deltabetaamp} simultaneously shifts $M_1^{(2)}(0)$ and $M_2^{(1)}(0)$ so they both coincide with the empirical Baldin sum rule value.
For the slope of $M_2^{(1)}$ at $Q^2=0$, we find the following contributions from $\pi N$ loops,  $\Delta$ exchange and $\pi\Delta$ loops, in units of $10^{-4}$~fm$^5$:
\begin{subequations}
\begin{align}
\left.\frac{\dd M_{2p}^{(1)} (Q^2)}{\dd Q^2}\right|_{Q^2=0}&= -3.92(38)\approx -1.47-2.18-0.26 ,  \\
\left.\frac{\dd M_{2n}^{(1)} (Q^2)}{\dd Q^2}\right|_{Q^2=0}&=-4.81(47) \approx -2.37-2.18-0.26 .
\end{align}
\end{subequations}
Interestingly, the slope does not show such a drastic cancellation between the $\pi N$-loop and
the $\Delta$-exchange contributions as one encounters in the generalized Baldin sum rule. Correspondingly,
the shape of the $M_2^{(1)}$ curve is not so much affected by the inclusion of the $g_M$ dipole
form factor, as one can see from Fig.~\ref{Fig:M14-orders} \{upper panel\} which
shows the individual contributions to the $Q^2$ dependence of $M_2^{(1)}(Q^2)$.

\subsection{\boldmath{$M^{(4)}_1(Q^2)$} --- the generalized fourth-order Baldin sum rule}

\begin{figure}[tbh]
\begin{center}
 \includegraphics[width=0.49\textwidth]{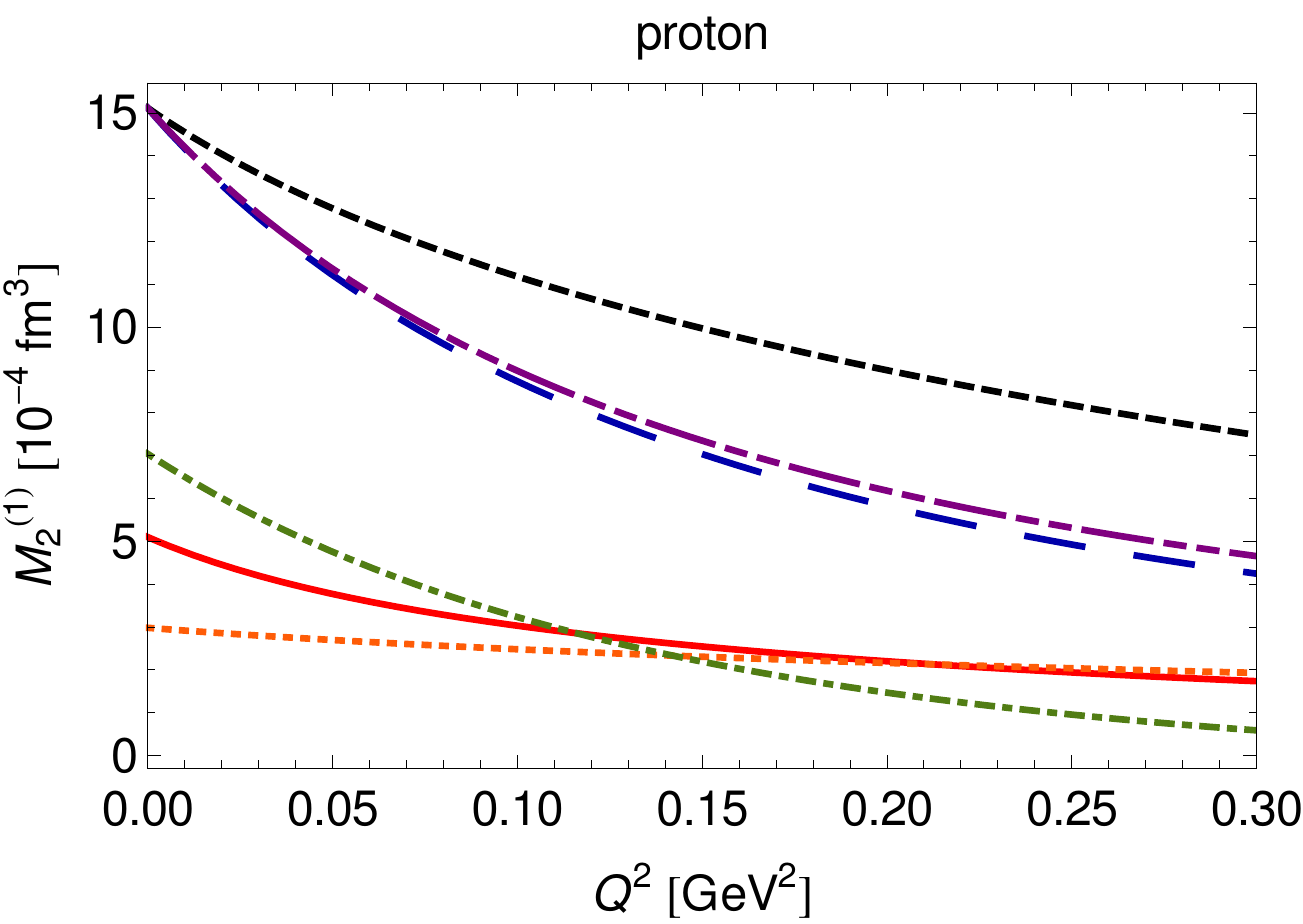}
  \includegraphics[width=0.49\textwidth]{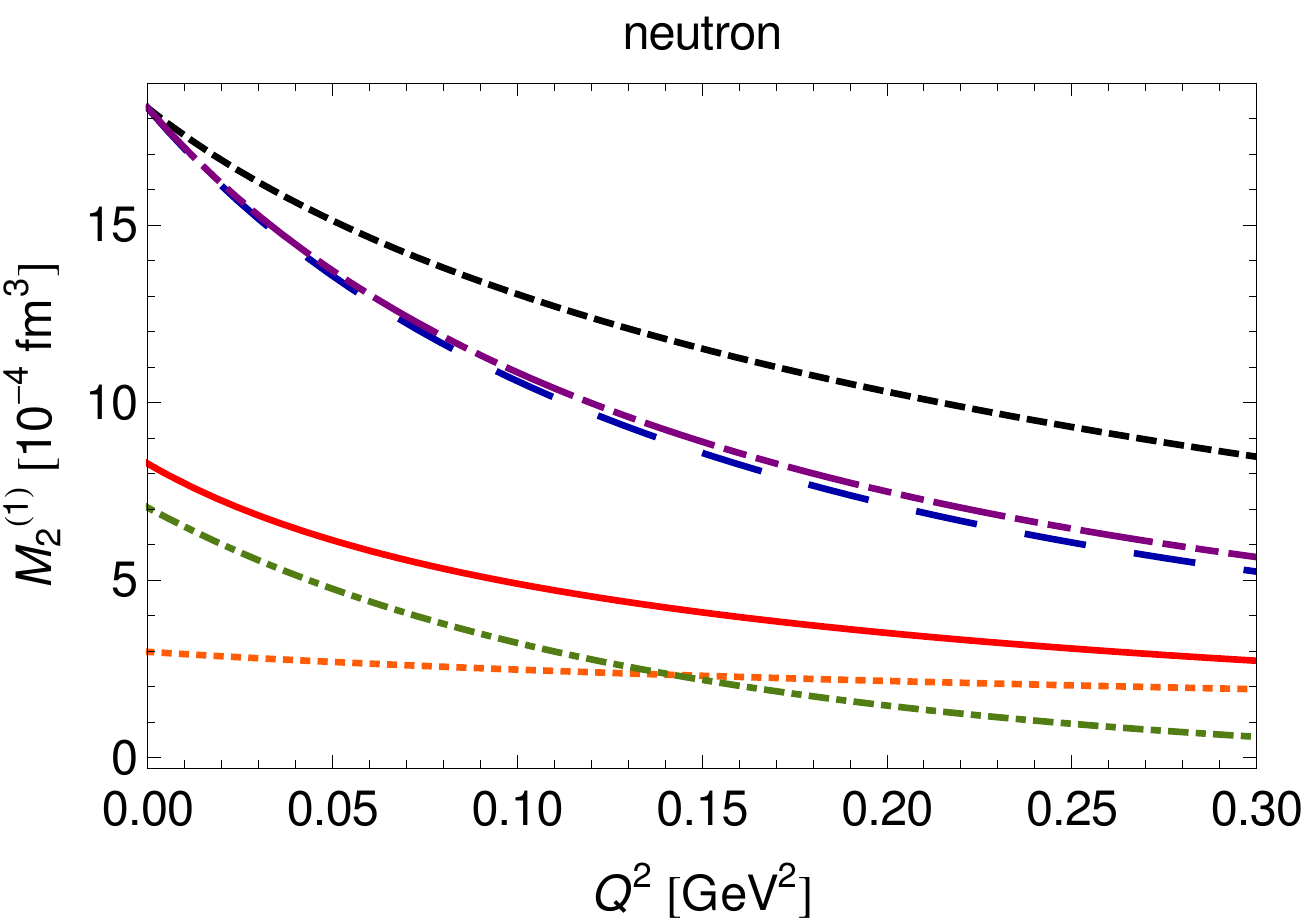}\\[0.2cm]
   \includegraphics[width=0.49\textwidth]{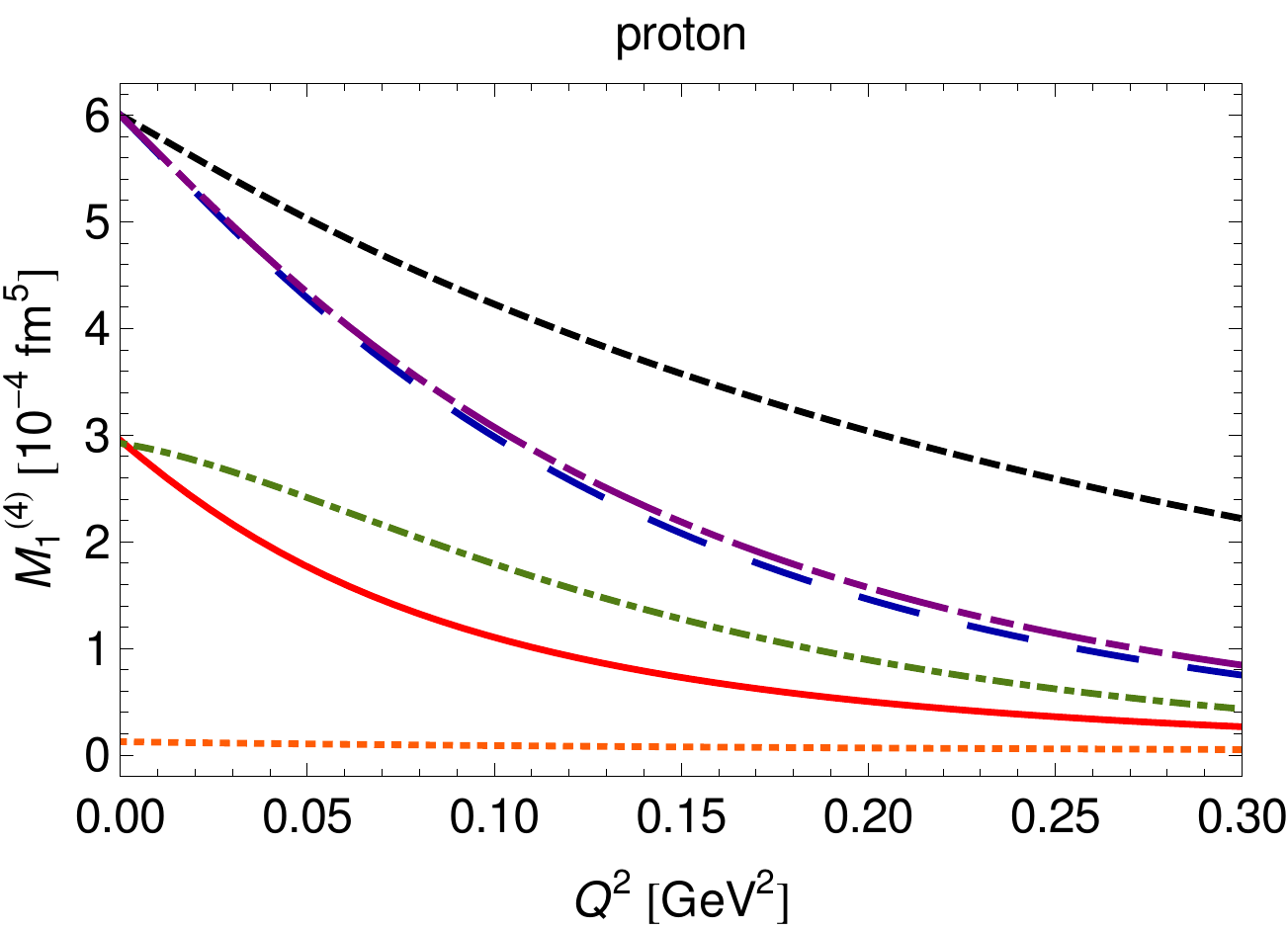}
    \includegraphics[width=0.49\textwidth]{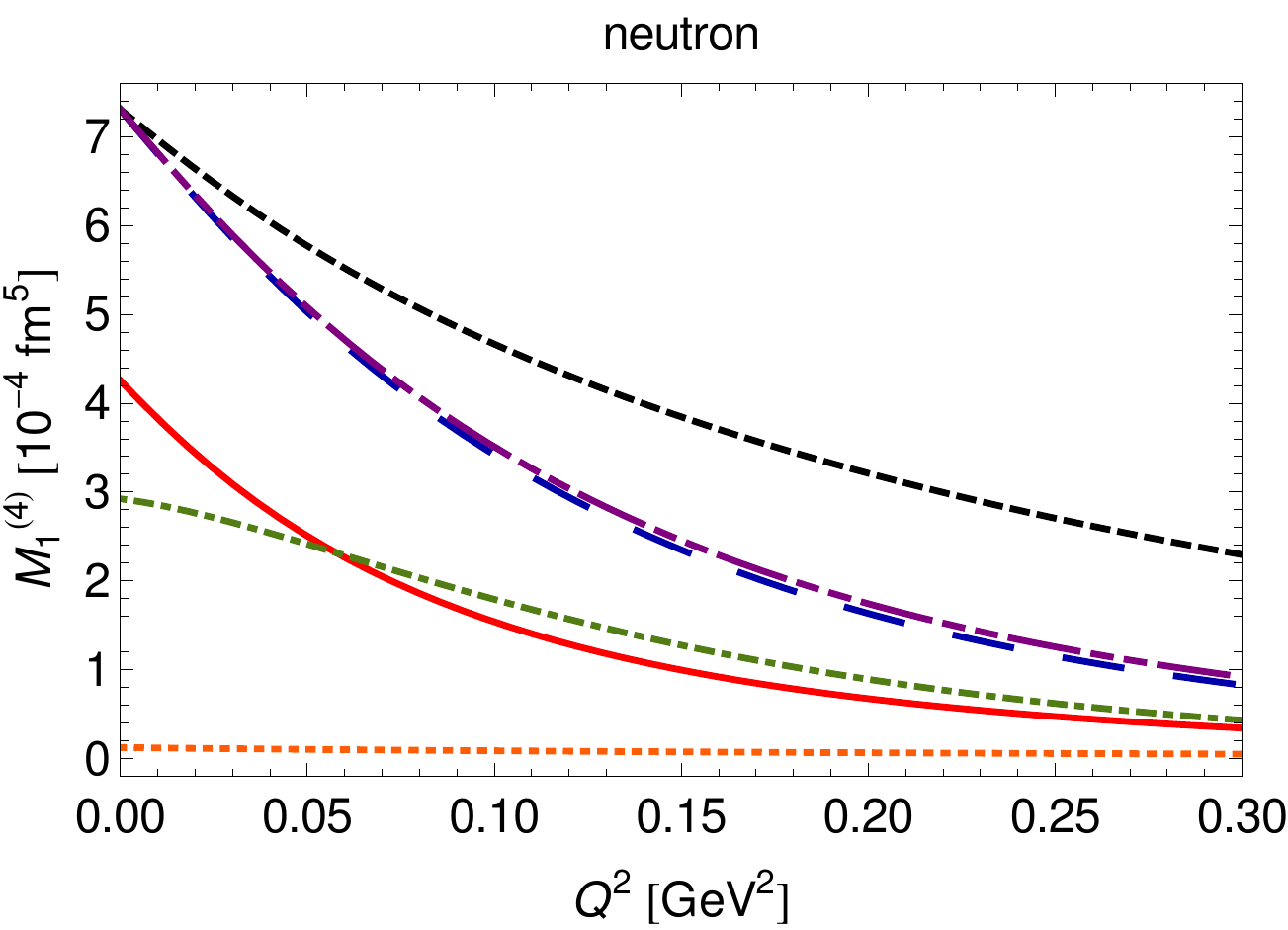}
\caption{\small{Contributions of the different orders to the chiral prediction of $M_2^{(1)}(Q^2)$ \{upper panel\} and $M_1^{(4)}(Q^2)$ \{lower panel\} for the proton (left) and neutron (right). Red solid line: $\pi N$-loop contribution, green dot-dashed line: $\Delta$-exchange contribution, orange dotted line: $\pi \Delta$-loop contribution, blue long-dashed line: total result, purple dot-dot-dashed line: total result without $g_C$ contribution, black short-dashed line: total result without $g_M$ dipole.}\label{Fig:M14-orders}}
\end{center}
\end{figure}

Let us now consider the fourth moment of the structure function $F_1(x,Q^2)$:
\bea
 M_1^{(4)}(Q^2)&=& 
\frac{1}{2 \pi^2} \, \int_{\nu_0}^{\infty}\, \mathrm{d}\nu \,\sqrt{1+\frac{Q^2}{\nu^{2}}}\, \frac{\sigma_T(\nu)}{\nu^{4} }\\
&=&\frac{32 \al M_N^3}{Q^8}\int_0^{x_0}\!\dd x\, x^3 \,F_1(x,Q^2) \nn.
\eea
In the real-photon limit, this moment is related to a linear combination of dispersive and quadrupole polarizabilities \cite{Guiasu:1978dz,Holstein:1999uu}, resulting in the fourth-order Baldin sum rule (see Ref.~\cite{Hagelstein:2015egb} for review):
\beq
M_1^{(4)}(0)=\alpha_{E1 \nu} + \beta_{M1 \nu} + \frac{1}{12} (\alpha_{E2} + \beta_{M2}) .
\eeq
Here we obtain the following NLO results for the proton and neutron (showing also the separate contributions from $\pi N$ loops,  $\Delta$ exchange, and $\pi\Delta$ loops), in units of $10^{-4}$~fm$^5$:
\begin{subequations}
\begin{align}
M_{1p}^{(4)}=6.00(59)\approx2.95+2.92+0.13, \\
M_{1n}^{(4)}=7.30(72) \approx 4.26+2.92+0.13.
\end{align}
\end{subequations}
For the slopes at $Q^2=0$, we find, in units of $10^{-4}$~fm$^7$:
\begin{subequations}
\begin{align}
\left.\frac{\dd M_{1p}^{(4)} (Q^2)}{\dd Q^2}\right|_{Q^2=0}&=-1.38(14)\approx -1.16-0.20-0.02  ,  \\
\left.\frac{\dd M_{1n}^{(4)} (Q^2)}{\dd Q^2}\right|_{Q^2=0}&=-1.96(19) \approx -1.73-0.20-0.02.
\end{align}
\end{subequations}
 The corresponding individual contributions to the $Q^2$ dependence
of $M_1^{(4)}(Q^2)$ are demonstrated in Fig.~\ref{Fig:M14-orders} \{lower panel\}.

In Fig.~\ref{Fig:M14plot} \{lower panel\}, we show our B$\chi$PT predictions compared to the MAID model predictions \cite{Drechsel:2000ct,Drechsel:1998hk,private-Lothar},  the  HB limit  of  the $\pi N$-loop  contribution \cite{Nevado:2007dd}, and empirical evaluations of the fourth-order Baldin sum rule \cite{Gryniuk:2015aa,Schroder:1977sn} for proton and neutron, respectively. Our NLO B$\chi$PT predictions are in good agreement with MAID, while the HB results fail to describe the decrease with growing $Q^2$.

\subsection{\boldmath{$\ol{T}_1(0,Q^2)$} --- the proton subtraction function}

The knowledge of the proton subtraction function $\ol{T}_1(0,Q^2)$ is needed
to evaluate the leading contribution of the nucleon structure to the (muonic-)hydrogen
Lamb shift, see Refs.~\cite{Hagelstein:2015egb,Pohl:2013yb,Carlson:2015jba} for reviews.
At very low momenta the non-Born part of the subtraction function is given by the magnetic dipole polarizability, $\ol{T}_1(0,Q^2)/Q^2=4\pi \beta_{M1}+\mathcal{O}(Q^2)$.
Since the Lamb-shift integrals
are weighted toward low $Q^2$, the low-momentum features of $\ol{T}_1(0,Q^2)$ 
have a more pronounced effect. 
In particular, the uncertainty in the (empirical) extraction of $\beta_{M1}$ contributes the
bulk of the theoretical uncertainty in Ref.~\cite{Birse:2012eb}. At the same time, the slope
of this function at $Q^2=0$ could potentially be important, and the different models and
mechanisms could lead to rather different values of that slope. 

\begin{figure}[bt]
\begin{center}
\epsfig{file=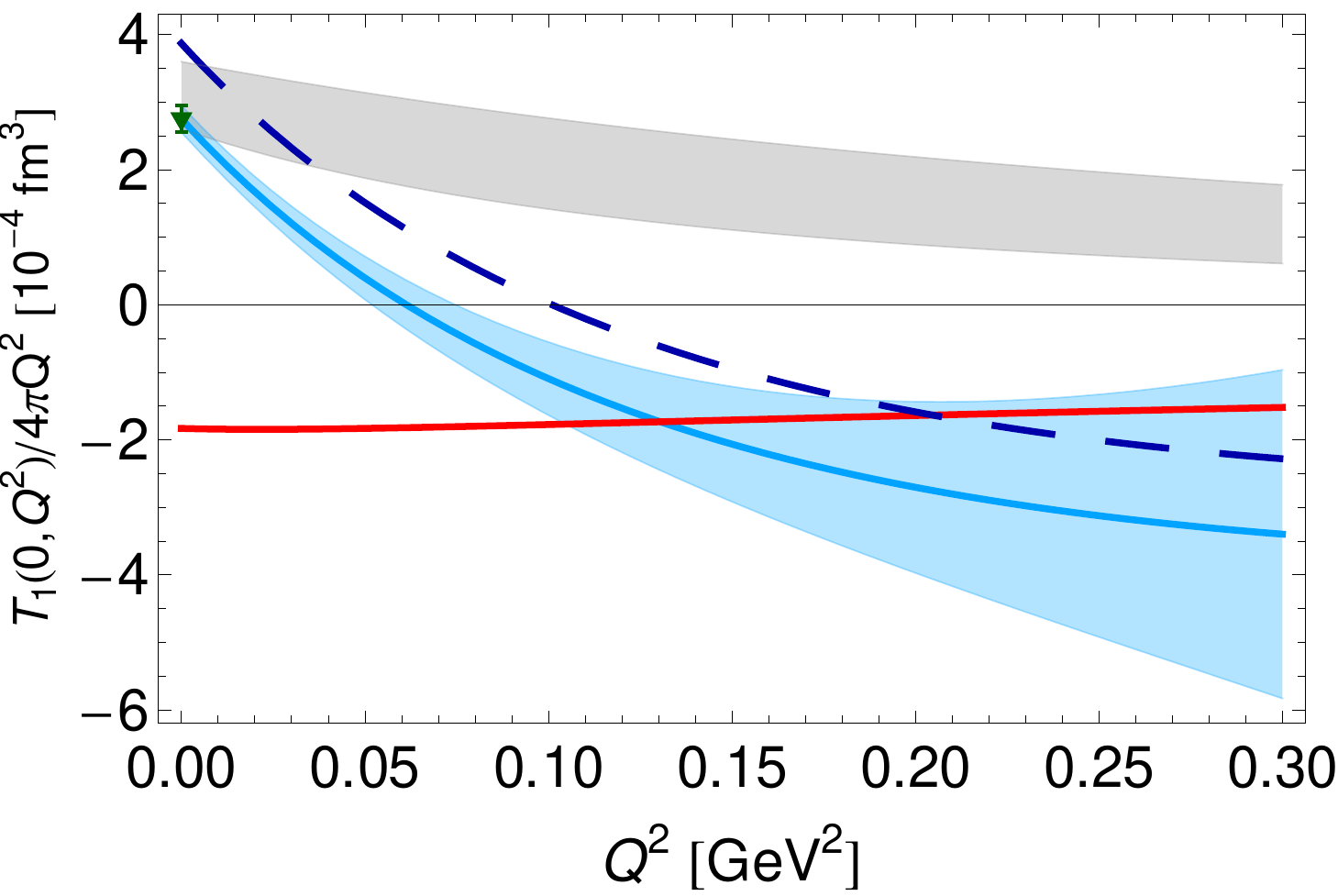,width=0.49\textwidth,angle=0}
\epsfig{file=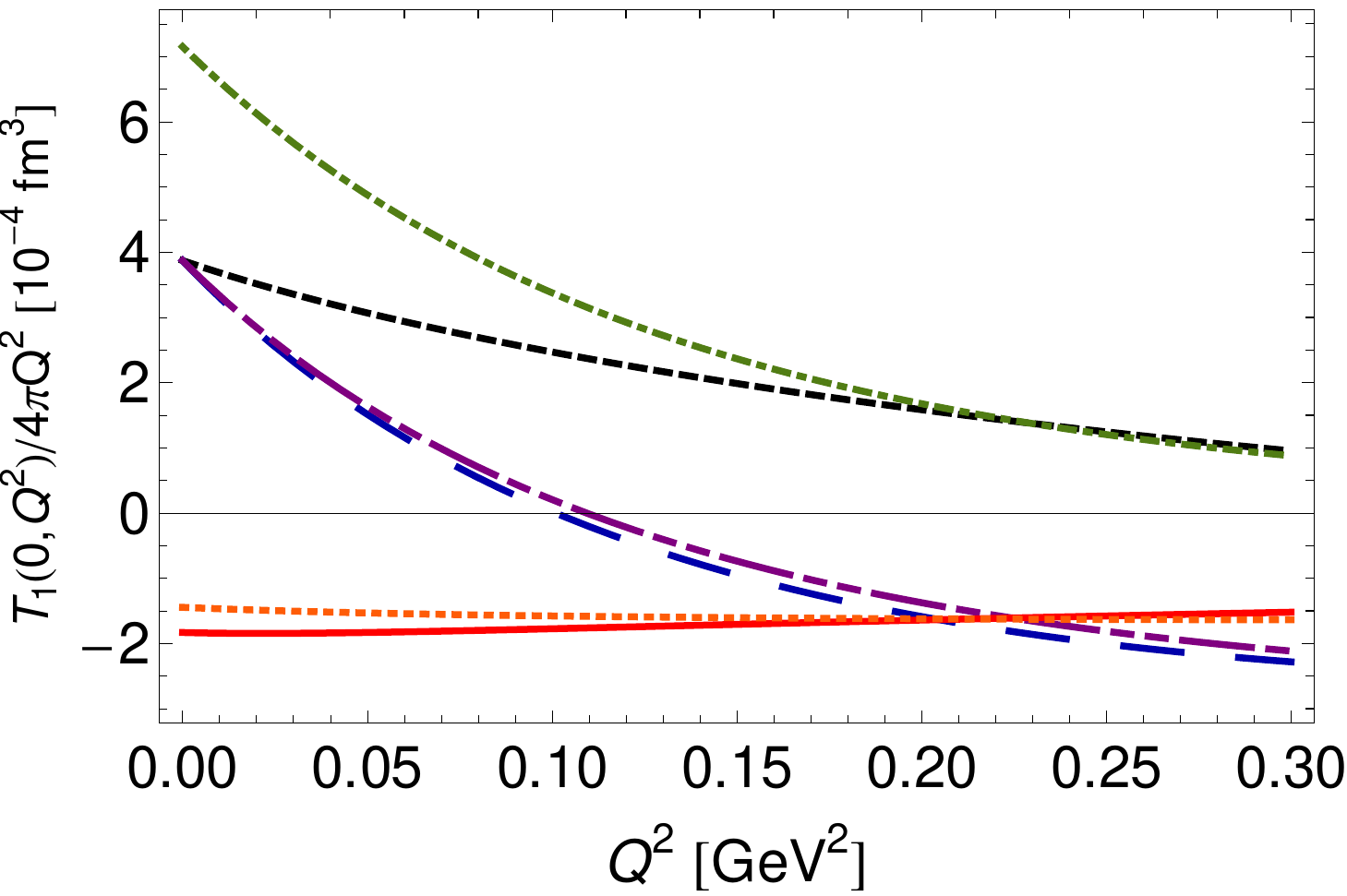,width=0.49\textwidth,angle=0}
\caption{\small{Left panel: The low-$Q^2$ behavior of the non-Born piece of  subtraction function $ \ol{T}_1(0,Q^2)/4\pi Q^2$ for the proton. The result of this work, including the $\delta\beta$ contribution, is shown by the blue solid line, with the blue band representing the uncertainty due to higher-order effects. The blue long-dashed line corresponds to the NLO B$\chi$PT prediction (i.e., without the $\delta\beta$ term).
At the real-photon point, we show the value of $\beta_{M1p}=(2.75\pm 0.2)\times 10^{-4}~\text{fm}^3$ (dark green triangle) resulting from the fit to the Baldin sum rule described in \secref{PowerCounting}. The red solid curve corresponds to the B$\chi$PT $\pi N$-loop contribution only, the gray band is the HB$\chi$PT evaluation~\cite{Birse:2012eb}. Right panel: Contributions of the different orders to the chiral prediction of $\ol{T}_1(0,Q^2)/4\pi Q^2$ for the proton. Red solid line: $\pi N$-loop contribution, green dot-dashed line: $\Delta$-exchange contribution, orange dotted line: $\pi \Delta$-loop contribution, blue long-dashed line: total result, purple dot-dot-dashed line: total result without $g_C$ contribution, black short-dashed line: total result without $g_M$ dipole.
}
}
\label{fig:subtraction}
\end{center}
\end{figure}

This is illustrated in Fig.~\ref{fig:subtraction} \{left panel\}, which shows the low-$Q^2$ behavior of $\ol{T}_1(0,Q^2)/4\pi Q^2$.
One can see that both $\beta_{M1}$ and the slope change significantly when one adds the $\Delta$ contributions ($\Delta$-exchange
and the $\pi\Delta$ loops contribution), cf.\ Fig.~\ref{fig:subtraction} \{right panel\}.
The resulting curve in the left panel is compared with the HB$\chi$PT evaluation of Ref.~\cite{Birse:2012eb}, showing an appreciable disagreement in the slope,
with the $Q^2$ dependence of the two curves being noticeably different.
Our NLO prediction of the slope at $Q^2=0$ is given by (in units of $10^{-4}$~fm$^5$):
\begin{align}
\frac{1}{8\pi}\left.\frac{\dd^2  \ol{T}_1(0,Q^2)}{\dd (Q^2)^2}\right|_{Q^2=0}&=-2.33(23)\approx -0.06-2.18-0.10.
\end{align}
Extracting the slope
of the subtraction function experimentally should in principle be possible through dilepton electroproduction as proposed in Ref.~\cite{Pauk:2020gjv}. It remains to be seen
whether such a measurement is feasible in the near future.

\section{Summary and conclusions}\label{Sec:Summary}

We have completed the NLO calculation of the
unpolarized VVCS amplitudes in SU(2) B$\chi$PT, with explicit $\Delta(1232)$. We have calculated the non-Born amplitudes, which at this order come out as a parameter-free prediction of B$\chi$PT. We have provided the theoretical uncertainty of these predictions due to higher-order effects, as well as an explicit illustration  
of such effects due to the inclusion of a low-energy constant from N$^2$LO [$\mathcal{O}(p^4)]$. The obtained VVCS amplitudes are used to examine several notable combinations
of the (generalized) polarizabilities that are expressed through the moments of the nucleon
structure functions, i.e.: $M_1^{(2)}(Q^2)$ --- the generalized Baldin sum rule, $M_1^{(4)}(Q^2)$ ---
the generalized fourth-order Baldin sum rule, $\alpha_L(Q^2)$ --- the longitudinal polarizability,
and $M_2^{(1)}(Q^2)$ --- the first moment of the structure function $F_2(x,Q^2)$.
The dispersion relations between the VVCS amplitudes and the tree-level photoabsorption cross sections served as a cross-check of these calculations. 

These results can be compared with the dispersive evaluations using the  empirical parametrization of the nucleon structure functions. The biggest discrepancy is 
observed for the low-$Q$ behavior of the generalized Baldin sum rule, calling
for a future revision of the low-momentum behavior of the empirical
parametrization of the structure function $F_1(x,Q^2)$.

Concerning the $\Delta(1232)$ contribution, we have seen that it plays an important role in transverse quantities, whereas in the longitudinal quantities, such as the longitudinal polarizability $\alpha_L$, its role is negligible. We have studied a modification of the magnetic $\gamma N \to \Delta$ coupling $g_M$  which 
incorporates the effects of vector-meson dominance 
[cf.\ \Eqref{modifiedgm}]and it turned out to be important in some cases, even at low $Q^2$.
This emphasises the importance of the VMD-type of effects in the $\gamma^* N \to \Delta$ transition form factor.
Strictly speaking, it needs to be included within the $\chi$PT in a more systematic manner, either
by an explicit inclusion of the vector mesons, or a resummation of the $\pi\pi$ rescattering. It would be
interesting to implement (one of) these systematic extensions of $\chi$PT in the future calculations. 
We have also considered the effect of 
the Coulomb (C2) $\gamma^* N \to \Delta$ transition, described by the coupling $g_C$. However, we find that it has generally a small effect in the unpolarized moments considered here.

We have obtained an NLO prediction for the proton subtraction function $\ol{T}_1(0,Q^2)$, which cannot be deduced from dispersion relations. This is an important step towards a systematic
improvement of the LO $\chi$PT evaluation~\cite{Alarcon:2013cba} of the proton-polarizability contribution
to the muonic-hydrogen Lamb shift.

\section*{Acknowledgements}

We thank Lothar Tiator and Marc Vanderhaeghen for helpful discussions. This work is supported by the Deutsche Forschungsgemeinschaft (DFG) through the
Collaborative Research Center [The Low-Energy Frontier of the Standard Model (SFB 1044)]. J.M.A. acknowledges support from the Community of Madrid through the ``Programa de atracci\'on de talento investigador 2017 (Modalidad 1)'', and the Spanish MECD grants FPA2016-77313-P. F.H. gratefully acknowledges financial support from the Swiss National Science Foundation.

\appendix
\small
\section{Photoabsorption cross sections}\label{App:CrossSections}

The forward CS amplitude can, up to the subtraction function, be
reconstructed from the total photoabsorption cross sections through the dispersion relations in \Eqref{genDRs}. Therefore, we use tree-level cross sections to verify our NLO calculation of the non-Born VVCS amplitudes.

\begin{figure*}[tbh]
\begin{center}
\epsfig{file=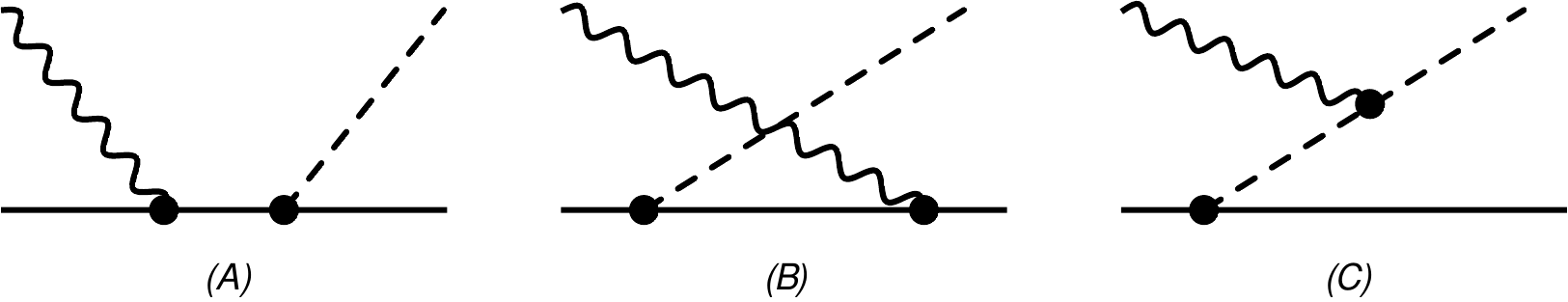,width=0.59\textwidth,angle=0} 
\caption{\small{Contribution of the $\ga^\ast N \to  \pi N$ channel to photoabsorption on the nucleon. \label{Fig:DiagsOp}}}
\end{center}
\end{figure*}

At LO, we need to consider the $\pi N$-production channel. Following Refs.~\cite{Holstein:2005db,Lensky:2009uv} and performing a chiral rotation to cancel exactly the Kroll-Ruderman term at this order, only the  tree-level diagrams shown in Fig.~\ref{Fig:DiagsOp} contribute, which are gauge invariant by themselves. Analytical expressions for $\sigma_T(\nu,Q^2)$ and $\sigma_L(\nu,Q^2)$ can be found in Ref.~\cite{Alarcon:2013cba}. The cross sections in the real-photon limit can be found in Refs.~\cite{Lensky:2009uv,Holstein:2005db}. We checked that the non-Born VVCS amplitudes at LO, the left-hand side of \Eqref{genDRs}, are reproduced by the right-hand side of the same equation when the tree-level $\pi N$-production photoabsorption cross sections are inserted. 

\begin{figure*}[tbh]
\begin{center}
\epsfig{file=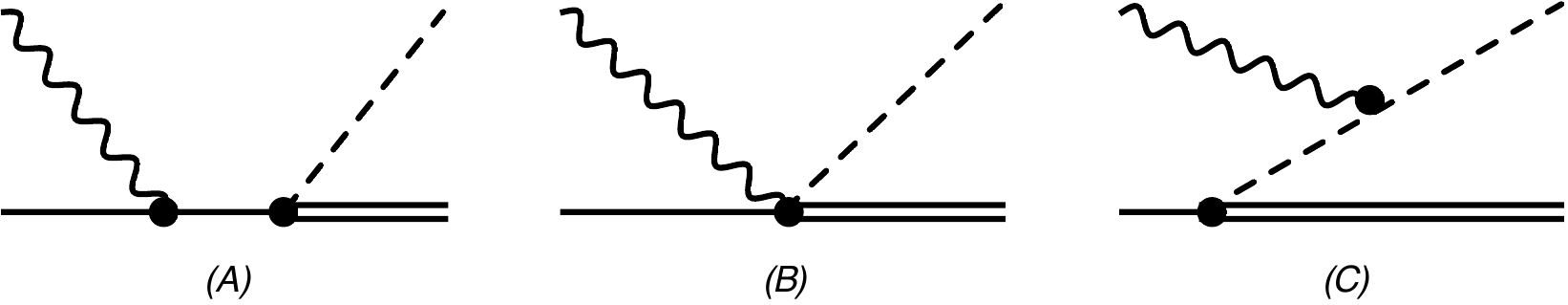,width=0.59\textwidth,angle=0} 
\caption{\small{Contribution of the $\ga^\ast N \to  \pi \Delta$ channel to photoabsorption on the nucleon. \label{Fig:DiagsDeltaCS}}}
\end{center}
\end{figure*}

\begin{figure}[tbh]
    \centering 
  \includegraphics[width=0.15\textwidth]{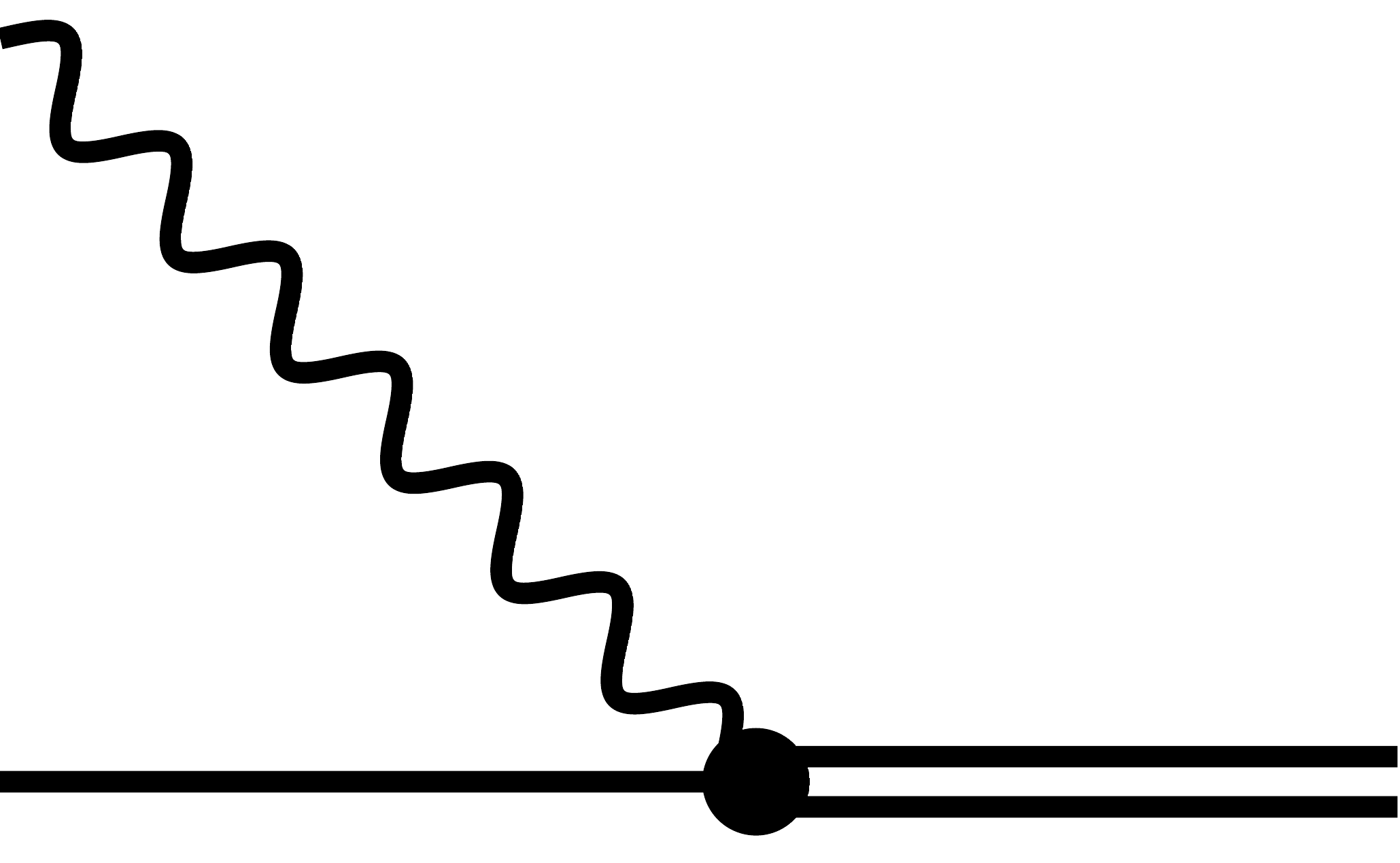}
\caption{\small{Contribution of the $\ga^\ast N \to  \Delta$ channel to photoabsorption on the nucleon.\label{fig:CrossSectionDeltaProd}}}
\end{figure}

\begin{figure*}[tbh]
\begin{center} 
\includegraphics[width=0.9\textwidth]{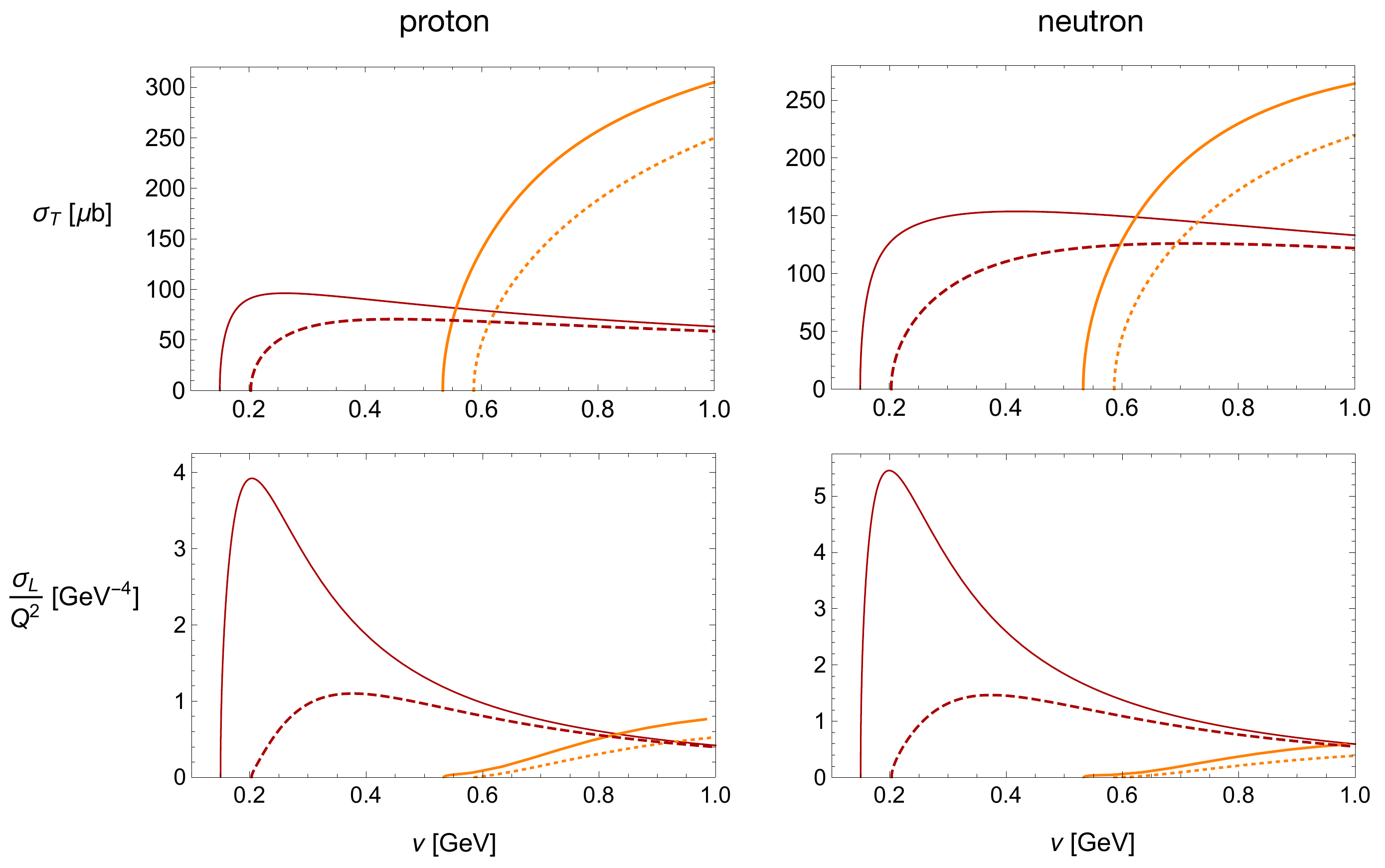}
\caption{\small{Photoabsorption cross sections for $\pi N$ (red) and $\pi \Delta$ production (orange) with $Q^2=0$ (solid) and $Q^2=0.1$ GeV$^2$ (dashed for $\pi N$ and dotted for $\pi \Delta$ channel). \label{Fig:SummaryCrossSections}}}
\end{center}
\end{figure*}

Besides the $\pi N$ production, we calculated the tree-level $\pi \Delta$-production and $ \Delta$-production photoabsorption cross sections, see Figs.~\ref{Fig:DiagsDeltaCS} and \ref{fig:CrossSectionDeltaProd}. Because of the worse high-energy behavior of the $\pi \Delta$-production cross sections, cf.\  Fig.~\ref{Fig:SummaryCrossSections}, the dispersion relations require further subtractions for a reconstruction of the $\pi \Delta$-loop contribution to the VVCS amplitudes. However, we could use these cross sections
to verify higher-order terms in the expansion of the VVCS amplitudes
in powers of small $\nu$.

The $\Delta$-production cross sections are related to the tree-level $\Delta$-exchange shown in Fig.~\ref{DeltaExchange}. The threshold for production of the $\Delta(1232)$-resonance is at lab-frame photon energies of:
\beq
\nu_\Delta=\frac{M_\Delta^2-M_N^2+Q^2}{2M_N}.
\eeq
Therefore, the $\Delta$-production cross sections contain to the following Dirac's $\de$-function: $\delta(\nu-\nu_\Delta)$. The
explicit form of these cross sections is given by:
\begin{subequations}
\bea
\si_T(\nu,Q^2)&=&\frac{4\pi^2 \al}{2M_NM_+^2\vert \vec{q}\,\vert}\Bigg\{g_M^2 \vert \vec{q}\, \vert ^2 (\nu +M_+)+\frac{g_E^2\, (\nu -\varDelta ) \left(M_N\nu -Q^2\right)^2}{M_N^2}\\
&&+\frac{g_C^2 \,Q^4 s (\nu -\varDelta )}{M_N^2 M_\Delta^2}-\frac{g_M g_E\, \vert \vec{q}\, \vert ^2 \left(M_N\nu -Q^2\right)}{M_N}+\frac{g_M g_C\, \vert \vec{q}\, \vert ^2 Q^2}{M_N}\nn\\
&&+\frac{2 g_E g_C \,Q^2 \left(M_N\nu -Q^2\right) [-M_\Delta(M_N+\nu)+s]}{M_N^2 M_\Delta}\Bigg\}\delta\!\left(\nu-\nu_\Delta\right),\nn\\
\si_L(\nu,Q^2)&=&\frac{4\pi^2 \al}{2M_N^3M_+^2\vert \vec{q}\,\vert}\Bigg\{g_E^2(\nu-\varDelta)\left[M_N^2 \vert \vec{q}\, \vert^2-(Q^2-M_N\nu)^2\right]\\
&&+\frac{g_C^2 Q^2(\nu-\varDelta)(M_N^2 \vert \vec{q}\, \vert^2-Q^2 s)}{M_\Delta^2}\nn\\
&&-\frac{2 g_E g_C \,Q^2 \left(M_N\nu -Q^2\right) [s-M_\Delta(M_N+\nu)]}{ M_\Delta}\Bigg\}\delta\!\left(\nu-\nu_\Delta\right),\nn
\eea
\end{subequations}
with $\varDelta=M_\Delta - M_N$, $M_+=M_\Delta + M_N$, and the Mandelstam variable $s=M_N^2+2M_N \nu-Q^2$.
Analytical expressions for the unpolarized structure functions can be constructed with the help of \Eqref{VVCSunitarity}, with the flux factor $K(\nu,Q^2)=\sqrt{\nu^2+Q^2}$.

It is important to note that the above cross sections only describe the $\Delta$-pole contributions to the tree-level $\Delta$ exchange. In general, the VVCS amplitudes described by the $\Delta$-exchange diagram in Fig.~\ref{DeltaExchange} can be split as follows \cite{Hagelstein:2018bdi}:
\begin{subequations}
\bea
\ol T_1^{\Delta\text{-exch.}}(\nu,Q^2)&=&\ol T_1^{\Delta\text{-exch.}}(0,Q^2)+T_1^{\Delta\text{-pole}}(\nu,Q^2)+\widetilde T_1^{\Delta\text{-exch.}}(\nu,Q^2),\qquad\;\\
\ol T_2^{\Delta\text{-exch.}}(\nu,Q^2)&=&T_2^{\Delta\text{-pole}}(\nu,Q^2)+\widetilde T_2^{\Delta\text{-exch.}}(\nu,Q^2).
\eea
\end{subequations}
$\ol T_1^{\Delta\text{-exch.}}(0,Q^2)$ is the usual subtraction function:
\bea
\ol T_1^{\Delta\text{-exch.}}(0,Q^2)&=&\frac{4\pi \al Q^4}{M_\Delta M_+ \omega_+}\left[\frac{g_M^2}{Q^2}-\frac{g_E^2 \varDelta}{M_N^2 M_+}-\frac{g_C^2 \varDelta \left(M_N^2-Q^2\right)}{M_N^2 M_\Delta^2 M_+}+\frac{g_M g_E}{M_N M_+}+\frac{g_M g_C}{M_N M_+}\right.\eqlab{T1su}\\
&&+\left.\frac{2 g_E g_C \left(M_N\varDelta +Q^2\right)}{M_N^2 M_\Delta M_+}\right].\nn
\eea
with  $\omega_\pm=(M_\Delta^2-M_N^2\pm Q^2)/2M_\Delta$. $T_i^{\Delta\text{-pole}}$ are the $\Delta$-pole contributions that feature a pole at the $\Delta(1232)$-production threshold, and thus, are proportional to:
\beq
\frac{1}{[s-M_\Delta^2][u-M_\Delta^2]}=\frac{1}{4M_N^2}\frac{1}{\nu_\Delta^2-\nu^2}, \eqlab{poleStruc}
\eeq
 where $s$ and $u$ are the usual Mandelstam variables. $\widetilde T_i^{\Delta\text{-exch.}}$ are the ($\Delta$-)non-pole terms in which the pole has canceled out \cite{Hagelstein:2018bdi}:
 \begin{subequations}
 \bea
\widetilde T_1^{\Delta\text{-exch.}}(\nu,Q^2)&=&-\frac{4\pi \al  \nu^2}{M_N M_+^2}\left(g_M^2+g_E^2-g_M g_E\right),\eqlab{T1nonpole}\\
\widetilde T_2^{\Delta\text{-exch.}}(\nu,Q^2)&=&-\frac{4\pi \al Q^2}{M_N M_+^2}\left(g_M^2+g_E^2-g_M g_E+\frac{g_C^2 Q^2}{M_\Delta^2}\right).\eqlab{T2nonpole}
\eea
\end{subequations}
 To describe the non-pole terms in Eqs.~(\ref{eq:T1nonpole}) and (\ref{eq:T2nonpole}) within the standard dispersive framework, \Eqref{genDRs}, we define the auxiliary structure functions:
 \begin{subequations}
\bea
\widetilde F_1(x,Q^2)&=&\frac{M_N x}{8 \pi \al} \,\widetilde T_1^{\Delta\text{-exch.}}(x,Q^2)\, \delta (x),\\
\widetilde F_2(x,Q^2)&=&\frac{Q^2}{16 \pi \al M_N} \,\widetilde T_2^{\Delta\text{-exch.}}(x,Q^2)\, \delta (x).
\eea
\end{subequations}

\section{Polarizabilities at $Q^2=0$}\label{App:PolarizabilitiesAll}

In this section, we give analytical expressions for the polarizabilities and their slopes at $Q^2=0$. In particular, we give the HB expansion of the $\pi N$-loop contributions and the $\Delta$-exchange contributions. The complete expressions, also for the $\pi \Delta$-loop contributions, can be found in the {\it Supplemental Material}.

\subsection{$\pi N$-loop  contribution}

Here, we give analytical expressions for the $\pi N$-loop contributions to the proton and neutron polarizabilities, expanded
in powers of $\mu = m_\pi/M_N$, viz., the HB expansion. Note that we
choose to expand here to a high order in $\mu$, the strict
HB expansion would only retain the leading term in an analogous
NLO calculation. 
\begin{itemize}
    \item Polarizabilities at $Q^2=0$:
\bea
\alpha_{E1p}+\beta_{M1p} &=&
  \frac{e^2 g_A^2}{96\pi^3 f_\pi^2\, m_\pi} \left\{  \frac{11 \pi }{8} + 6  (3 \log \mu +4)\mu - \frac{1521 \pi \mu^2}{64} \right.\nn\\
  &&\left.-\frac{ (210 \log \mu + 29)\mu^3 }{3  } +\dots\right\},\\
\alpha_{E1n}+\beta_{M1n} &=&
 \frac{e^2 g_A^2}{96\pi^3 f_\pi^2\, m_\pi} \left\{  \frac{11 \pi}{8} + \frac{ (12 \log \mu +1)\mu}{2} - \frac{117 \pi \mu^2}{64} +\frac{ 7 \mu^3 }{3 } +\dots\right\}, \\
 \alpha_{Lp} &=&
 \frac{e^2 g_A^2}{1440 \pi^3 f_\pi^2\,  m_\pi^3}\left\{ \frac{93 \pi}{32} - \frac{89 \mu}{2} + \frac{18231 \pi \mu^2}{256} +10 (44+ 51 \log\mu )\mu^3  \right. \nn\\
&&\left.-\frac{1880805 \pi \mu^4}{4096} - 3\left(356 \log\mu -\frac{129}{10}\right)\mu^5 +\dots  \right\},\\
\alpha_{Ln} &= &
 \frac{e^2 g_A^2}{1440 \pi^3 f_\pi^2\,  m_\pi^3}\left\{ \frac{93 \pi}{32} - \frac{35}{2} \mu + \frac{4095 \pi \mu^2}{256} + \frac{1}{2}(11+ 120 \log\mu )\mu^3 \right. \nn\\
 &&\left.-\frac{80085 \pi \mu^4}{4096}+ \frac{141\mu^5}{5} +\dots  \right\},
 \eea
 \bea
M_{1p}^{(4)} (0)&=&\frac{e^2 g_A^2}{720 \pi^3 f_\pi^2 m_\pi^3}\left\{\frac{81 \pi}{32}-56 \mu+\frac{29145\pi\mu^2}{256}+\frac{3}{4}\left(1501+1380 \log \mu \right)\mu^3\right. \nn\\
&&\left.-\frac{4670925\pi \mu^4}{4096}-\frac{96}{5}\left(13+160\log \mu\right)\mu^5 +\dots\right\},\\
M_{1n}^{(4)}(0) &= &\frac{e^2 g_A^2}{720 \pi^3 f_\pi^2 m_\pi^3}\left\{\frac{81\pi}{32}-28\mu+\frac{6525\pi\mu^2}{256}+3\left(1+30\log \mu\right)\mu^3 \right.\nn\\
&&\left.-\frac{113925\pi\mu^4}{4096} +\frac{192\mu^5}{5} +\dots\right\}.
\eea
\item Slopes of polarizabilities at $Q^2=0$:
\begingroup
\allowdisplaybreaks
\bea
\frac{\dd\big(\alpha_{E1p}+\beta_{M1p}\big) (0)}{\dd Q^2}
&=& \frac{e^2 g_A^2}{480 \pi^3 f_\pi^2\,  m_\pi^3} \left\{  \frac{3 \pi}{16} -18 \mu + \frac{6477 \pi \mu^2}{128} + \left( \frac{1339}{2} + 550 \log\mu\right)\!\mu^3 \right. \qquad \nn\\* 
&& \left. -\frac{1366515 \pi \mu^4}{2048}-7\left( \frac{313}{10} + 270 \log\mu  \right)\mu^5 +\dots  \right\},\\
\frac{\dd\big(\alpha_{E1n}+\beta_{M1n}\big) (0)}{\dd Q^2}&=&
 \frac{e^2 g_A^2}{1440 \pi^3 f_\pi^2\,  m_\pi^3} \left\{  \frac{9 \pi}{16} -\frac{89 \mu}{2} + \frac{5535 \pi \mu^2}{128} + \left( 1+ 150 \log\mu\right)\mu^3  \right.\nn\\*
 &&\left.-\frac{92265  \pi \mu^4}{2048} +  \frac{1209 \mu^5}{20} +\dots  \right\},\\
\frac{\dd\alpha_{Lp} (0)}{\dd Q^2}&=& \frac{e^2 g_A^2}{1440 \pi^3 f_\pi^2  m_\pi^5}\left\{-\frac{621\pi}{896}+\frac{13\mu}{14}+\frac{4995\pi \mu^2}{1024} - \frac{669\mu^3}{7}\right. \nn\\*
&&\left.+\frac{2517315\pi\mu^4}{16384}+ \left( \frac{34407}{35} + 1116 \log \mu\right)\mu^5+\dots \right\},\\
\frac{\dd\alpha_{Ln} (0)}{\dd Q^2}&=&\frac{e^2 g_A^2}{1440 \pi^3 f_\pi^2  m_\pi^5}\left\{-\frac{621\pi}{896}-\frac{55\mu}{14}+\frac{3195\pi\mu^2}{1024}-\frac{207\mu^3}{7}\right.\nn\\*
&&\left.+\frac{456915\pi\mu^4}{16384}+\frac{18}{35}(29+210 \log \mu)\mu^5+\dots \right\},\\
\frac{\dd M_{2p}^{(1)} (0)}{\dd Q^2}&=&\frac{e^2 g_A^2}{480\pi^3 f_\pi^2 m_\pi^3}\left\{-\frac{17\pi}{32}+\frac{9\mu}{2}-\frac{399\pi\mu^2}{256}+\frac{1}{3}\left(197+90\log \mu\right)\mu^3\right.\nn\\*
&&\left.-\frac{246015\pi\mu^4}{4096}-\frac{1}{5}\left(199+990\log \mu\right)\mu^5+\dots \right\},\\
\frac{\dd M_{2n}^{(1)} (0)}{\dd Q^2}&=&\frac{e^2 g_A^2}{480\pi^3 f_\pi^2 m_\pi^3}\left\{-\frac{17\pi}{32}-2\mu+\frac{705\pi\mu^2}{256}+\frac{1}{6}\left(1+60\log \mu\right)\mu^3\right.\nn\\*
&&\left.-\frac{12255\pi\mu^4}{4096}+\frac{79\mu^5}{20}+\dots \right\},\\
\frac{\dd M_{1p}^{(4)}(0) }{\dd Q^2}&=&\frac{e^2g_A^2}{10080\pi^3f_\pi^2m_\pi^5}\left\{\frac{225\pi}{128}-229\mu+\frac{495555\pi\mu^2}{1024}-7542\mu^3\right.\nn\\*
&&+\frac{217523775\pi\mu^4}{16384}+3\left(38771+37408\log \mu\right)\mu^5+\dots \Big\},\\
\frac{\dd M_{1n}^{(4)}(0) }{\dd Q^2}&=&\frac{e^2g_A^2}{2520\pi^3f_\pi^2m_\pi^5}\left\{\frac{225\pi}{512}-\frac{209\mu}{4}+\frac{228795\pi\mu^2}{4096}-\frac{1653\mu^3}{4}\right.\nn\\*
&&+\frac{21341775\pi\mu^4}{65536}+3\left(3+364\log \mu\right)\mu^5+\dots \Big\}.
\eea
\endgroup
\end{itemize}
\subsection{$\Delta$-exchange contribution}\seclab{DeltaTreePolarizabilities}

Here, we give analytical expressions for the  tree-level $\Delta$ exchange contributions, cf.\ Fig.~\ref{DeltaExchange}, to the nucleon polarizabilities and their slopes at $Q^2=0$. Note that the $\Delta$-exchange  contributes equally to proton and neutron polarizabilities. 
Recall that for the magnetic $\gamma^* N \Delta $ coupling we introduced a dipole form factor to mimic vector-meson dominance: $g_M \rightarrow g_M/(1+Q^2/\Lambda^2)^2$. 
\begin{itemize}
    \item Polarizabilities at $Q^2=0$:
\bea
\al_{E1}&=&-\frac{e^2 g_E^2}{2 \pi  M_+^3},\\
\be_{M1}&=&\frac{e^2 g_M^2}{2 \pi    M_+^2}\frac{1}{\varDelta},\\
\al_L&=&\frac{e^2 M_\Delta^2}{\pi M_+^3}\left(\frac{ g_E^2}{\varDelta  M_NM_+^2}-\frac{g_C^2}{2 M_\Delta^4}+\frac{ g_E g_C}{M_N M_\Delta^2 M_+}\right),\\
M_1^{(4)}(0)&=&\frac{e^2 M_N}{\pi M_+^3 \varDelta} \left(\frac{g_M^2}{\varDelta^2}+\frac{g_E^2}{M_+^2}-\frac{g_E g_M}{\varDelta M_+}\right).
\eea
\item Slopes of polarizabilities at $Q^2=0$:
\bea
\frac{\dd \left[\alpha_{E1}+\beta_{M1}\right](0)}{\dd Q^2}&=&-\frac{e^2}{\pi M_+^2}\left(\frac{g_M^2}{\varDelta^2}\left[\frac{1}{M_+}-\frac{1}{2\varDelta}\right]+\frac{2}{ \Lambda^2 }\frac{g_M^2}{\varDelta}\right.+\frac{g_M g_E}{M_N}\left[\frac{1}{4\varDelta^2}-\frac{1}{\varDelta M_+}\right.\nn\\
&&\left.\left.+\frac{1}{4M_+^2}\right] -\frac{g_E^2}{4M_NM_+}\left[\frac{1}{\varDelta}-\frac{5}{M_+}\right]-\frac{ g_M g_C}{2  \varDelta  M_NM_+}+\frac{g_E g_C}{M_NM_+^2}\right),\qquad\\
\frac{\dd \al_L(0)}{\dd Q^2}&=&\frac{e^2 M_\Delta^3}{\pi \varDelta M_+^4}\left(\frac{2g_E^2}{\varDelta^2 M_+^2}\left[\frac{2}{M_\Delta}-\frac{1}{M_N}\right]-\frac{g_C^2}{ M_\Delta^4}\left[\frac{1}{M_N}-\frac{3}{2M_\Delta}\right]\right.  \nn\\
&&+\left.\frac{g_E g_C}{\varDelta M_\Delta^2 M_+}\left[\frac{5}{M_\Delta}-\frac{3}{M_N}\right]\right),\\
\frac{\dd M_2^{(1)} (0)}{\dd Q^2}&=&\frac{e^2 }{\pi M_+^3}\left(-\frac{g_M^2}{2\varDelta^2}-\frac{2g_M^2 M_+}{\Lambda^2 \varDelta}+\frac{g_M g_E}{2\varDelta M_N }+\frac{g_E^2}{2\varDelta M_+}+\frac{g_M g_C}{2\varDelta M_N}\right.\nn\\
&&\left.-\frac{g_C^2}{2M_\Delta^2}\right),\\
\frac{\dd M_1^{(4)} (0)}{\dd Q^2}&=&\frac{e^2 M_N}{\pi \varDelta^2 M_+^5}\left(\frac{g_M^2}{\varDelta}\left[2-\frac{5M_+}{\varDelta}+\frac{M_+^2}{\varDelta^2}\right]-\frac{4M_+^2}{\varDelta}\frac{g_M^2}{\Lambda^2}+g_M g_E \left[\frac{6}{\varDelta}\right.\right.\nn\\
&&\left.\left.-\frac{M_+}{\varDelta^2}-\frac{1}{M_+}\right]+\frac{2M_+ g_M g_E}{\Lambda^2}+g_E^2\left[\frac{1}{\varDelta}-\frac{7}{M_+}\right]+\frac{2g_M g_C}{\varDelta}\right.\nn\\
&&\left.-\frac{4g_E g_C}{M_+}\right).
\eea
\end{itemize}


  \small
\bibliography{lowQ}

\end{document}